\documentclass{ieeeaccess}
\usepackage{cite}
\usepackage{amsmath,amssymb,amsfonts}
\usepackage{algorithmic}
\usepackage{graphicx}
\usepackage{textcomp}
\usepackage{float}
\usepackage{soul}
\usepackage{changepage}
\usepackage{xcolor}
\usepackage{graphicx}
\usepackage{subcaption}
\usepackage{multirow}
\usepackage{xcolor}
\usepackage{booktabs}
\usepackage{makecell}
\usepackage{soul}
\usepackage{comment}
\definecolor{blue}{RGB}{0,0,255}
\usepackage{CJKutf8}
\def\BibTeX{{\rm B\kern-.05em{\sc i\kern-.025em b}\kern-.08em
T\kern-.1667em\lower.7ex\hbox{E}\kern-.125emX}}
\usepackage{xcolor}
\usepackage{pict2e}
\newsavebox{\ORCIDlogo}
\savebox{\ORCIDlogo}{%
\setlength{\unitlength}{\dimexpr 1em/256\relax}%
\begin{picture}(256,256)%
  \color[HTML]{A6CE39}\put(128,128){\circle*{256}}%
  \color{white}%
  \put(78.6,199.2){\circle*{20}}%
  \moveto(70.9,176,9)\lineto(86.3,176,9)\lineto(86.3,69.8)\lineto(70.9,69.8)%
  \closepath\fillpath%
  \moveto(108.9,176.9)\lineto(150.5,176.9)%
  \curveto(190.1,176.9)(207.5,148.6)(207.5 ,123.3)%
  \curveto(207.5,95,8)(186,69.7)(150.7,69.7)%
  \lineto(108.9,69.7)%
  \closepath\fillpath%
  \color[HTML]{A6CE39}%
  \moveto(124.3,83.6)\lineto(148.8,83.6)%
  \curveto(183.7,83.6)(191.7,110.1)(191.7,123.3)%
  \curveto(191.7,144.8)(178,163)(148,163)%
  \lineto(124.3,163)%
  \closepath\fillpath%
\end{picture}%
}
\newcommand\orcidicon[1]{\href{https://orcid.org/#1}{\usebox{\ORCIDlogo}}}
\usepackage[hidelinks]{hyperref}

\usepackage{hyperref} %<--- Load after everything else
\IEEEoverridecommandlockouts
\usepackage{cite}
\usepackage{amsmath,amssymb,amsfonts}
\usepackage{algorithmic}
\usepackage{graphicx}
\usepackage{textcomp}
\usepackage{xcolor}
\def\BibTeX{{\rm B\kern-.05em{\sc i\kern-.025em b}\kern-.08em
    T\kern-.1667em\lower.7ex\hbox{E}\kern-.125emX}}
\begin{document}
\history{Date of publication xxxx 00, 0000, date of current version xxxx 00, 0000.}
\doi{10.1109/ACCESS.2017.DOI}
\title{Computer-Aided Osteoporosis Diagnosis Using Transfer Learning with Enhanced Features from Stacked Deep Learning Modules}
%\title{Computer-Aided Osteoporosis Diagnosis Using ResNet50 Feature Extraction Enhanced by Stacked Deep Learning Modules}
\author{
\uppercase{Ayesha Siddiqua} \orcidicon{0009-0007-9893-1768}\authorrefmark{1},
%\uppercase{Fahmid Al farid \ \orcidicon{0009-0004-0120-8388} \authorrefmark{2}},
\uppercase{Rakibul Hasan}\orcidicon{0009-0005-3707-2344}\authorrefmark{3},
%\uppercase{sarina mansor\authorrefmark{2}},
\uppercase{Anichur Rahman} \authorrefmark{3},
\uppercase{abu saleh musa miah\ \orcidicon{0000-0002-1238-0464} \authorrefmark{1}(Member, IEEE)} 
}
\address[1]{Department of Computer Science and Engineering, Bangladesh Army University of Science and Technology, Saidpur, Nilphamari, Bangladesh}
%\address[2]{Faculty of Engineering, Multimedia University, Cyberjaya 63100, Malaysia}
\address[2]{Department of Computer Science and Engineering, Brac University, Merul Badda, Dhaka, Bangladesh} 
\address[3]{Department of Computer Science and Engineering, National Institute of Textile Engineering and Research (NITER), Constituent Institute of the University of Dhaka, Savar, Dhaka-1350, Bangladesh}

%\tfootnote{This work was supported by Multimedia University, Cyberjaya, Selangor, Malaysia, under Grant MMUI/230023.02.}
%\address[2]{School of Computer Science and Engineering,  
\markboth
{Author \headeretal: Preparation of Papers for IEEE TRANSACTIONS and JOURNALS}
{Author \headeretal: Preparation of Papers for IEEE TRANSACTIONS and JOURNALS}
\corresp{Corresponding authors: Abu Saleh Musa Miah (abusalehcse.ru@gmail.com)}% and Sarina Mansor (sarina.mansor@mmu.edu.my)}

\begin{abstract}
Knee osteoporosis weakens the bone tissue in the knee joint lead to raising the risk of fractures and early detection of the osteoporosis using  X-ray image  allows for timely intervention and better patient outcomes. Many researsher have been working to devlope various medical image system but a few researcher have been working to develop knee Osteoporosis using manual radiology evaluation and traditional machine learning with hand-crafted features. However, these methods often face challenges in performance accuracy and efficiency due to reliance on manual feature extraction and subjective interpretation by radiologists. In this research, we propose a computer-aided diagnosis system for knee osteoporosis using transfer learning combined with stacked feature enhancement deep learning blocks to overcome the challenges. Initially, knee X-ray images undergo preprocessing, followed by feature extraction using a pre-trained Convolutional Neural Network (CNN). Further, these extracted features are then refined and enhanced through five sequential Conv-RELU-MaxPooling blocks, each designed to capture and refine hierarchical feature representations progressively. The Conv2D layers detect low-level features, while the ReLU activations introduce non-linearity, enabling the network to learn complex patterns. Also, the MaxPooling layers down-sample the features, retaining only the most important spatial information. This sequential stacking of blocks allows the model to capture increasingly complex, high-level features related to bone structure, joint deformation, and osteoporotic markers. The enhanced features are then passed through a classification module to differentiate between healthy and osteoporotic knee conditions. Most importantly, extensive experiments with three individual datasets and a large combined dataset demonstrated that our model generated 97.32\%, 98.24\%, 97.27\%, and 98.00\% accuracy for OKX Kaggle Binary, KXO-Mendely Multi-Class. OKX Kaggle Multiclass and combined dataset, respectively, which is around 2\% higher than the existing method. Additionally,  the advantages of the proposed system were demonstrated by its high-performance accuracy. This approach demonstrates a robust and efficient method for accurately classifying knee osteoporosis from X-ray images.
\end{abstract}
\titlepgskip=-15pt
\maketitle

\section{Introduction}
Osteoporosis is a debilitating bone disease characterized by a progressive decrease in bone density and deterioration in bone quality \cite{yu2022osteoporosis}, which significantly increases the risk of fractures and deformities. This condition often develops silently over many years, typically remaining undetected until a significant fracture occurs. The delayed diagnosis of osteoporosis poses serious challenges, highlighting the critical importance of early detection and intervention for effective disease management. Traditional diagnostic methods primarily focus on bone mineral density (BMD) measurements and imaging techniques \cite{de2022osteoporosis}, which can be labour-intensive and require a high level of specialized expertise. Knee X-rays, while primarily used to diagnose osteoarthritis, can also offer valuable insights into bone density and structural changes, which may be indicative of osteoporosis. Despite their potential, manual analysis of knee X-rays to detect osteoporosis presents significant challenges. The complexity of interpreting X-ray images and the subtlety of bone density changes require expertise that may not always be available in clinical settings. 
Osteoporosis, characterized by reduced bone density and quality, increases the risk of fractures and deformities \cite{kanis2013european}. This condition often progresses silently, remaining undetected until a fracture occurs \cite{carlson2019review}. Traditional diagnostic methods, such as bone mineral density (BMD) measurements and imaging, are labor-intensive and require specialized expertise \cite{haseltine2021bone}. In addition to being frequently used to diagnose osteoarthritis, knee X-rays can reveal information on bone density and structural alterations associated with osteoporosis \cite{chen2023state}. However, interpreting these X-rays is challenging due to the subtle nature of bone density changes, requiring expert-level interpretation that is not always available in clinical settings \cite{alzubaidi2024comprehensive}.

\subsection{Challenges in Current Knee Osteoporosis Disease Detection Work}
Recent advancements in medical imaging and artificial intelligence (AI) have significantly improved diagnostic accuracy and efficiency. Machine learning methods, including traditional image processing techniques, have been utilized to detect osteoporosis \cite{anam2021osteoporosis,albuquerque2023osteoporosis,eyedisease_miah_2024}, a condition characterized by decreased bone density and increased fracture risk. Early detection of osteoporosis is crucial for preventing fractures, as the condition often progresses silently until a fracture occurs \cite{kanis2013european}. Several studies have employed classical machine learning techniques, often using features extracted from X-ray images to predict osteoporosis or osteopenia \cite{lee2020exploration,xue2023osteoporosis}. These methods usually rely on manual feature extraction and simple classifiers. However, traditional machine learning methods often struggle with accuracy, especially when dealing with complex, high-dimensional data like medical images.

\vspace{2mm}
Deep Learning (CNN)-based Approaches Convolutional Neural Networks (CNNs), a type of deep learning model, have revolutionized image analysis tasks in various domains, including medical imaging \cite{salehi2023study,khan2022multinet}. CNNs excel at learning hierarchical patterns and features from raw image data, making them particularly effective for image-based classification tasks such as osteoporosis detection from knee X-rays. Unlike traditional machine learning, CNNs can automatically learn relevant features from images, eliminating the need for manual feature extraction. Deep learning, particularly CNNs, has demonstrated efficiency in detecting osteoporosis from knee X-rays, even with relatively limited datasets \cite{chawla2022prediction}. For example, a study using a Kaggle dataset of 646 knee X-ray images showed that grayscale images outperformed RGB images in terms of accuracy and training efficiency \cite{abubakar2022transfer}. Another study employed a modified U-Net with attention units to segment bone regions in X-ray and DEXA images, achieving an accuracy of 88\%. However, this method faced challenges such as a restricted dataset and high computational demands, limiting its practical use \cite{nazia2020diagnosis}.
Further advancements have refined CNN-based methods to improve diagnostic accuracy. For instance, Chen et al. \cite{chen2023glcm} proposed a pre-processing method that combined a gray-level co-occurrence matrix (GLCM) with a fuzzy broad learning system (FBLS) to classify knee osteopenia and osteoporosis, achieving an accuracy of 87.09\%. Although this approach outperformed other CNN models, it was limited by a small dataset of only 239 images, affecting its generalizability and robustness.

\vspace{2mm}
Transfer learning has emerged as a powerful approach to improve the performance of convolutional neural networks (CNNs) in specialized tasks, such as osteoporosis detection, especially when working with limited annotated datasets \cite{dhanagopal2024channel}. Transfer learning involves adapting pre-trained models—those trained on large, general datasets such as ImageNet—for more specific tasks, such as medical image classification. This method allows models to leverage the knowledge learned from large-scale datasets and fine-tune them for specialized medical imaging tasks, improving performance with less data. Several studies have applied transfer learning in osteoporosis detection. For example, Sarhan et al. \cite{sarhan2024knee} employed transfer learning with models such as VGG-19 and ResNet-50 to improve classification accuracy. Their method achieved 92\% accuracy for multiclass classification and 97.5\% accuracy for binary classification, addressing some scalability and clinical applicability issues. However, the studies were still constrained by small datasets and the lack of multimodal data integration. In another study, Sarhan et al. also employed data augmentation techniques to enhance the diversity of the dataset, thereby improving the generalizability of the model. Despite these advancements, data scarcity and trust in AI models remain significant challenges. Moreover, the need for explainable AI models was emphasized to improve transparency and clinical adoption \cite{alzubaidi2024comprehensive}. Moreover, recent work has proposed the use of CNNs combined with transfer learning for an improved classification between osteoporosis and osteopenia based on Bone Mass Density values. This approach has shown reduced false positive rates and improved diagnostic accuracy \cite{ramesh2024multi}. However, models such as KONet, which employ weighted ensemble methods for knee osteoporosis detection from X-rays, achieved 97\% accuracy, but their complexity and small dataset size still limit their real-world applicability \cite{rasool2024konet}.
While the studies above \cite{abubakar2022transfer,nazia2020diagnosis,dhar2023challenges,chen2023glcm,sarhan2024knee,alzubaidi2024comprehensive} have contributed to the development of knee osteoporosis detection systems, several challenges remain. These include lower performance, limited generalizability, and overfitting due to small datasets. The primary cause of these issues is the insufficient extraction of effective hierarchical features. 
\subsection{Motivation}
Existing models for the detection of knee osteoporosis using X-rays face challenges related to the size, generalizability, and overfitting of the dataset, limiting their clinical applicability. The lack of effective hierarchical feature extraction impedes the accurate detection of subtle bone density changes. To address these issues, we aim to develop a more generalized detection system by leveraging enhanced features from stacked deep learning models, incorporating larger and more diverse datasets, and improving model interpretability for real-world use. Our goal is to create a more robust and accurate diagnostic tool that can aid healthcare professionals in early osteoporosis detection, ultimately improving patient outcomes.
\subsection{The Goal and Scope of the Study} To tackle these challenges, we propose the use of Enhanced Features derived from Stacked Deep Learning models to develop a generalized knee osteoporosis detection system. This approach leverages larger datasets to assist doctors in improving interpretability for practical clinical applications. The key contributions of the proposed system to enhance osteoporosis diagnosis using knee X-rays are outlined below: 
\begin{itemize}
    \item \textbf{Development of a Computer-Aided Model for Osteoporosis Diagnosis:} We developed a machine learning model to classify knee X-ray images for osteoporosis diagnosis, incorporating advanced deep learning techniques to improve diagnostic accuracy.
    \item \textbf{Investigation of Preprocessing Methods:} We explored various image preprocessing techniques, such as resizing, normalization, and data augmentation, to optimize the input images for model training and improve classification performance.
    \item \textbf{Integration of Transfer Learning:} By leveraging a pre-trained convolutional neural network (CNN) based on ResNet50, we utilized transfer learning to extract robust initial features from knee X-ray images, capitalizing on pre-learned weights from extensive datasets.
    
    \item \textbf{Stacked Feature Enhancement Deep Learning Blocks:} We introduced a series of five sequential Conv-RELU-MaxPooling blocks to further enhance the extracted features. These blocks capture progressively complex and hierarchical feature representations, crucial for distinguishing between healthy and osteoporotic knees.
    
    \item \textbf{Improved Generalization and Accuracy:} Extensive experiments with three individual datasets and a large combined dataset demonstrated that our model generated 97.32\%, 98.24\%, 97.27\%, and 98.00\% accuracy for OKX Kaggle Binary, KXO-Mendely Multi-Class. OKX Kaggle Multiclass and combined dataset, respectively, which is around 2\% higher than the existing method. The high performance demonstrated the superiority of the proposed model, which integrates transfer learning with stacked feature enhancement blocks and achieves higher performance accuracy. This combination enabled the model to generalize effectively across diverse input data, reducing overfitting and increasing diagnostic accuracy in clinical settings. 
\end{itemize}

\section{Literature Review}
\label{sec: Literature Review}
Osteoporosis is a skeletal disorder characterized by a decrease in bone density and deterioration of bone quality, which significantly increases the risk of fractures \cite{kanis2013european}. The condition often develops silently over many years, remaining asymptomatic until a fracture occurs, which can lead to severe complications and a decline in quality of life \cite{bliuc2015risk}. The World Health Organization (WHO) defines osteoporosis as a bone mineral density (BMD) that is 2.5 standard deviations or more below the mean value of a young adult reference population \cite{kanis2013european}. This condition poses a significant public health challenge, particularly in aging populations, necessitating effective screening and diagnostic strategies.
Ramesh et al.\cite{ramesh2024multi} highlights the role of sequential analysis techniques in advancing medical data analysis for early disease detection, particularly for osteoporosis and osteopenia, which primarily affect postmenopausal women. The research proposes a novel deep convolutional neural network-based sequential classifier to analyze health datasets and improve classification accuracy. By leveragingBone Mass Density values, the approach effectively distinguishes osteoporosis and osteopenia patients from healthy individuals. The proposed method achieves higher classification accuracy and reduces false positives, offering significant improvements in early disease prediction and healthcare outcomes.
\subsection{Traditional Diagnostic Methods}
\label{sec: Traditional Diagnostic Methods}
The gold standard for diagnosing osteoporosis has traditionally been dual-energy X-ray absorptiometry (DEXA), which measures BMD \cite{morgan2017quality}. DEXA is recognized for its accuracy and reliability; however, its limitations include the requirement for specialized equipment and trained personnel, which may not be available in all clinical settings \cite{cauley2011defining}. Furthermore, reliance on BMD alone may not fully capture the complexity of bone quality and fracture risk, leading to a need for complementary diagnostic approaches \cite{cosman2014clinician}.
\subsection{Alternative Diagnostic Approaches}
\label{sec: Alternative Diagnostic Approaches}
Recent studies have explored the potential of knee X-rays as a diagnostic tool for osteoporosis. While primarily used for evaluating osteoarthritis, knee X-rays can provide insights into bone density and structural changes associated with osteoporosis \cite{zhang2021osteoporosis}. These radiographs capture fine details of the cortical and trabecular bone patterns, allowing clinicians to observe early signs of bone thinning and other morphological changes associated with osteoporosis. Despite their promise, interpreting knee X-rays for osteoporosis diagnosis remains challenging due to the subtle nature of osteoporotic changes. Unlike fractures, which are visually apparent, the reduction in bone density is often gradual and less conspicuous, demanding a high level of radiological expertise and precision. %X-Rays Orthopedics: Diagnosing Bone and Joint Issues. Need to cite this link.
 This complexity can lead to variability in diagnostic accuracy, as minor bone density reductions may be difficult to detect and assess consistently, particularly when subtle structural changes overlap with signs of osteoarthritis or other degenerative joint diseases. In addition, Singh et al. focused on the integration of attention mechanisms with CNN models to further improve osteoporosis detection from knee X-rays. Their study highlighted the role of attention layers in capturing critical features associated with bone density reduction. This approach represents an innovative advancement in utilizing deep learning \cite{hossain2023crime_miah,rahim2024enhanced_miah} for osteoporosis screening, potentially improving early detection rates \cite{singh2022systematic}.

\subsection{Machine Learning in Osteoporosis Diagnosis}
\label{sec: Machine Learning in Osteoporosis Diagnosis}
A study by, Yasaka et al. \cite{yasaka2020prediction} explores the use of deep learning to predict bone mineral density (BMD) of lumbar vertebrae from unenhanced abdominal CT images. A convolutional neural network (CNN) model was trained using axial CT images and BMD values from dual-energy X-ray absorptiometry (DXA) as references. The model was validated internally and externally with data from two institutions, showing a significant correlation between the predicted BMD (r = 0.852 for internal validation and r = 0.840 for external validation) and the DXA values. The model achieved high diagnostic performance with AUCs of 0.965 (internal) and 0.970 (external), demonstrating that deep learning can accurately predict BMD and aid in osteoporosis diagnosis using unenhanced CT images. \\
Nazia et al.\cite{nazia2020diagnosis} improved osteoporosis diagnosis by accurately measuring Bone Mineral Density (BMD) and T-score values from X-ray images, addressing the limitations of DEXA machines' availability in developing countries. A modified U-Net with an Attention unit was used for precise segmentation of bone regions from X-ray and DEXA images. A linear regression model calculated BMD and T-scores, classifying images as normal, osteopenia, or osteoporosis. The method achieved 88\% accuracy on two datasets (DEXSIT and XSITRAY).  The results were validated against clinical reports, demonstrating the potential of digitized X-rays for efficient and accurate osteoporosis detection.\\ 
Another study involved 1,449 patients by Fang \cite{fang2021opportunistic} leverages deep learning to diagnose primary osteoporosis using CT images. A fully automated system combined U-Net for vertebral segmentation and DenseNet-121 for bone mineral density (BMD) calculation, trained on 586 cases and tested on 863 cases. Testing datasets were divided by CT vendor diversity, achieving high segmentation accuracy (dice coefficients: 0.823, 0.786, 0.782) and strong correlation (r > 0.98) between automated and QCT-derived BMD values. The approach successfully identified osteoporosis, osteopenia, and normal BMD levels, demonstrating its effectiveness for automated osteoporosis diagnosis. \\ 
Kumar et al. \cite{kumar2023fuzzy} explored a cost-effective approach for early osteoporosis detection by categorizing knee X-ray images into normal, osteopenia, and osteoporosis classes. Using a dataset of 240 X-rays (37 normal, 154 osteopenia, and 49 osteoporotic cases), the study employed pre-trained CNN architectures—Inception v3, Xception, and ResNet18—alongside pre-processing techniques like normalization and data augmentation. A fuzzy rank-based ensemble model was utilized to enhance diagnostic accuracy, achieving a classification accuracy of 93.5\% and AUC values of 98.1 for normal, 97.9 for osteopenia, and 97.3 for osteoporosis. While the study demonstrated significant potential, limitations such as reliance on transfer learning and the need for annotated data were acknowledged. Future work aims to refine deep learning models, incorporate additional imaging modalities, and validate the method clinically for broader applicability. 
Recent studies have focused on developing machine learning models \cite{zobaed2020real_miah,kibria2020creation_miah} tailored for the classification of knee X-ray images to diagnose osteoporosis. A study by Wani et al. utilized convolutional neural networks (CNNs) for classifying knee X-rays into healthy and osteoporotic categories, achieving an accuracy of over 90\%. Their results suggest that automated image analysis can be a valuable tool in screening for osteoporosis, particularly in clinical settings with limited access to advanced imaging modalities \cite{wani2023osteoporosis}.

\subsection{Advancements in Medical Imaging and AI}
\label{sec: Advancements in Medical Imaging and AI}
Recent studies have emphasized the potential of deep learning techniques in enhancing the accuracy of osteoporosis diagnosis from knee X-ray images. A comprehensive study by 
develops a system for diagnosing osteoporosis from X-ray images using transfer learning in two phases. In Phase 1, a deep convolutional neural network (DCNN) is trained on a large, unlabeled dataset from orthopedic centers to capture general features. In Phase 2, the model is fine-tuned with a target dataset. The pre-processing includes noise reduction, contrast enhancement, and region of interest extraction. The model achieved an accuracy of 94.5\% on Dataset A and 91.5\% on Dataset B. With transfer learning, accuracy improved to 98.91\% for Dataset A and 96.61\% for Dataset B, demonstrating the model's effectiveness in diagnosing osteoporosis \cite{mohammed2023diagnosis}.
 A study by Abubakar et al.\cite{abubakar2022transfer} highlights the use of Deep Learning models (GoogLeNet, VGG-16, ResNet50) for classifying osteoporosis using knee X-ray images. The dataset was split into grayscale and RGB formats, with training times compared (e.g., GoogLeNet: 42 min for grayscale, 50 min for RGB). It evaluates performance using state-of-the-art neural network metrics, showcasing AI's potential in osteoporosis diagnosis.\\
Knee osteoporosis (KOP) is a condition characterized by low bone density and bone degradation, increasing fracture risk. Traditional knee radiography relies on expert interpretation, which can be prone to misclassification due to subtle image variations. To address this, Rasool et al. (2024) proposed KONet, a robust model using a weighted ensemble approach to accurately detect normal and osteoporotic knee conditions. Trained on the Kaggle dataset with 372 X-ray images, KONet achieved 97\% accuracy. The model, utilizing transfer learning with state-of-the-art CNNs, outperformed existing methods, demonstrating superior diagnostic performance \cite{rasool2024konet}.
A comparative analysis by Chen et al. evaluated various CNN architectures, including VGG16, InceptionV3, and Xception, to assess their performance in identifying osteoporotic features in knee X-ray images. Their research found that the InceptionV3 model, when fine-tuned on a knee X-ray dataset, achieved an accuracy of 91\%, showing the model's adaptability and precision in classifying images for osteoporosis diagnosis. These findings suggest that transfer learning using diverse CNN architectures can significantly enhance diagnostic accuracy in radiological applications \cite{khan2022accurate}. Table \ref{tab:Comparison_of_Related_Work} summarizes the key aspects of various studies and their methodologies.

\begin{table*}[!htp] 
\centering
\scriptsize
\caption{Comparison of Related Work} 
\label{tab:Comparison_of_Related_Work} 
\begin{tabular}{|p{1.5cm}|p{1cm}|p{2.5cm}|p{1.2cm}|p{1.1cm}|p{3.3cm}|p{3cm}|} 
\hline  
\textbf{Study} & \textbf{Year} & \textbf{Dataset} & \textbf{Image Type} & \textbf{No. of Images} & \textbf{Techniques} & \textbf{Accuracy} \\ 
\hline 
Nazi et al. \cite{nazia2020diagnosis} & 2020 & DEXSIT and XSITRAY & \begin{tabular}[c]{@{}l@{}} X-ray and\\ DEXA images\end{tabular} & - & U-Net & 88\% \\ 
\hline 
Yasaka et al. \cite{yasaka2020prediction} & 2020 & - & CT images & 99,900 & CNN model & \begin{tabular}[c]{@{}l@{}} AUC of 0.965 (internal), \\ 0.970 (external)\end{tabular} \\ 
\hline 
Fang et al. \cite{fang2021opportunistic} & 2021 & - & CT image & 1449 & U-Net and DenseNet-121 & - \\ 
\hline 
Abubakar et al. \cite{abubakar2022transfer} & 2022 & Kaggle Dataset & X-ray & 372 & GoogLeNet, VGG-16, ResNet50 & 90\% \\ 
\hline 
Kumar et al. \cite{kumar2023fuzzy} & 2023 & Mendeley Dataset & X-ray & 240 & Fuzzy Rank-based Ensemble Model & 93.5\% \\ 
\hline 
Wan et al. \cite{wani2023osteoporosis} & 2023 & \begin{tabular}[c]{@{}l@{}} BMD camp organized \\ by the Unani and \\ Panchkarma Hospital \end{tabular} & X-ray & 381 & \begin{tabular}[c]{@{}l@{}} Transfer learning using \\ AlexNet, VGG16, ResNet, VGG19 \end{tabular} & 91.10\% \\ 
\hline 
Mohammed et al. \cite{mohammed2023diagnosis} & 2023 & Kaggle dataset & X-ray & - & \begin{tabular}[c]{@{}l@{}} Transfer learning with \\ deep convolutional neural network \end{tabular} & \begin{tabular}[c]{@{}l@{}} 98.91\% for Dataset A \\ and 96.61\% for Dataset B \end{tabular} \\ 
\hline 
Chen et al. \cite{chen2023glcm} & 2023 & Mendeley dataset & X-ray & 239 & GLCM, FBLS, SMOTE & 87.09\% \\ 
\hline 
Sarhan et al. \cite{sarhan2024knee} & 2024 & Kaggle and Mendeley data & X-ray & 1947 & VGG-19 & \begin{tabular}[c]{@{}l@{}} 92.0\% for multiclass \\ and 97.5\% for binary \end{tabular} \\ 
\hline 
Rasool et al. \cite{rasool2024konet} & 2024 & Kaggle dataset & X-ray & 372 & KONet (weighted Ensemble Model) & 97\% \\ 
\hline 
Naguib et al. \cite{naguib2024new} & 2024 & Kaggle and Mendeley dataset & X-ray & 240 and 371 & Superfluity DL & \begin{tabular}[c]{@{}l@{}} 85.42\% for dataset1 \\ and 79.39\% for dataset2 \end{tabular} \\ 
\hline
\end{tabular} 
\end{table*}

For future work, it is essential to explore the computational resources required for training and deploying deep learning models, particularly for applications such as osteoporosis detection using X-rays. Understanding the resource demands will help assess the scalability and practicality of implementing these models in real-world clinical settings. Additionally, further investigation into the interpretability of deep learning algorithms is crucial. By improving the transparency of model decision-making processes, we can enhance trust in the predictions and outcomes, ultimately fostering better integration of these technologies into healthcare practices. This could also lead to advancements in model refinement and the development of more reliable diagnostic tools for osteoporosis.

\section{Dataset Description}
\label{sec:Dataset}
\begin{comment}
    also update the below content in datset with citation: We have used three publicly used dataset Dataset-1 \cite{}, Dataset-2 \cite{} and dataset-3 \cite{}, to evaluate our proposed model. Table \ref{tab:Description of the datasets} shows the number of images of each class, and the description of each dataset is given in the following subsections. The datasets used for this study comprise knee X-ray images categorized into two classes: 'Healthy' and 'Osteoporosis'. 
\end{comment}
Osteoporosis is the second most prevalent bone disorder after arthritis, affecting millions globally. While Dual Energy X-ray Absorptiometry (DXA) is the gold standard for diagnosis \cite{sangondimath2023dexa}, its high cost and limited accessibility create barriers to widespread use. Treatment of osteoporosis is also expensive, emphasizing the need for cost-effective and accessible detection systems. Researchers have been striving to develop early detection methods that are affordable and efficient, but these efforts are often constrained by the use of localized, proprietary datasets. We have used three publicly used datasets, OKX Kaggle Binary Dataset \cite{stevepython_osteoporosis}, KXO-Mendely Multi-Class Dataset\cite{majeed2021knee_dataset_mendeley}, and OKX Kaggle Multiclass Dataset \cite{gobara2022knee_dataset_kaggle}, to evaluate our proposed model. Table \ref{tab:Description of the datasets} shows the number of images of each class, and the description of each dataset is given in the following subsections \ref{tab:Description of the datasets}. The dataset was balanced, ensuring an equal number of images for each class to prevent bias during model training and evaluation. 

\begin{table*}
\centering
    \caption{Description of the datasets}
    \label{tab:Description of the datasets}
    \begin{tabular}{|p{2cm}| p{1cm}| p{2cm}| p{2cm}| p{1cm}| p{1cm}|p{1cm}| p{1cm}|} \hline
        Dataset& No. of Classes & Class Name for the Binary Classes &  Class Name for the Multi-Classes & No. of Images in Class 1 & No. of Images in Class 2 & No. of Images in Class 3& Total Sample \\ \hline
        OKX Kaggle Binary Dataset & 2 & Normal and Osteoporosis & Normal, Osteopenia and Osteoporosis & 186 & 186& - & 372 \\ \hline
        KXO-Mendely Multiclass Dataset  \cite{majeed2021knee_dataset_mendeley}& 3 &Normal and Osteoporosis & Normal, Osteopenia and Osteoporosis & 36 & 154 & 49 &  239 \\ \hline
        OKX Kaggle Multiclass Dataset& 3 &Normal and Osteoporosis & Normal, Osteopenia and Osteoporosis & 780 & 374 &793 & 1947 \\ \hline
        Combined Dataset& 3 &Normal and Osteoporosis & Osteoporosis & 1002 & 528 & 1028 &2030 \\ \hline
    \end{tabular}
\end{table*}

\subsection{Osteoporosis Knee X-ray Binary Dataset (OKX Kaggle Binary Dataset)}
\label{sec: Kaggle Dataset 1}
This dataset \cite{stevepython_osteoporosis} contains X-ray images of knees classified into different categories related to osteoporosis. It aims to aid in the diagnosis and research of knee osteoporosis by providing a variety of images that represent the condition. The dataset is valuable for training machine learning models \cite{miah2019eeg,miah2019motor} to identify and classify the severity of osteoporosis based on imaging data, facilitating early detection and intervention. The Osteoporosis Knee X-ray dataset contains images labeled as either normal or osteoporotic.
%Dataset Link - https://www.kaggle.com/datasets/stevepython/osteoporosis-knee-xray-dataset
\begin{figure}[htp]
    \centering
    \includegraphics[width=8cm]{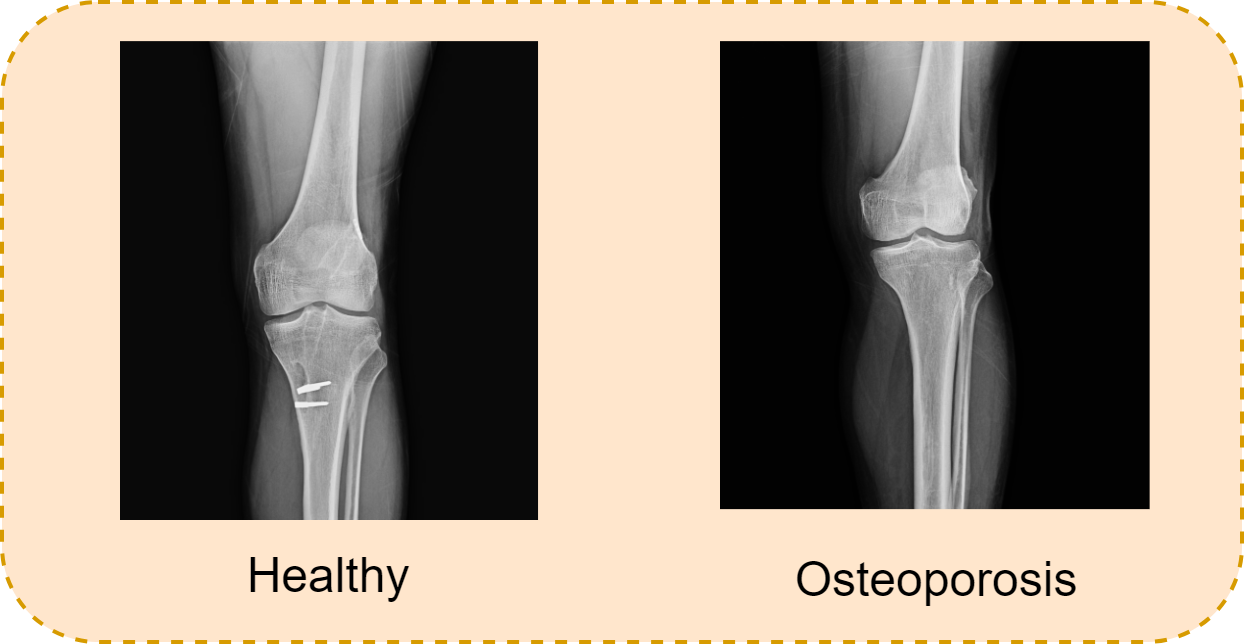}
    \caption{Visual Information of Knee X-rays: Healthy bone and affected by Osteoporosis from OKX Kaggle Binary Dataset}
    \label{fig:Sample Image}
\end{figure}
\subsection{Knee X-ray Osteoporosis Mendeley Dataset (KXO-Mendely Multi-Class Dataset)}
\label{sec: Mendeley Dataset}
%Age 40-75
%%%Majeed Wani, Insha ; Arora, Sakshi (2021), “Knee X-ray Osteoporosis Database”, Mendeley Data, V2, doi: 10.17632/fxjm8fb6mw.2
We also used a benchmark dataset which is publicly available in Mendely \cite{majeed2021knee_dataset_mendeley}. This dataset constructed with knee X-ray images for each participant, offering a detailed perspective on bone health. The primary objective of this dataset is to support the global research community in developing data-driven solutions for osteoporosis detection. By focusing on the relationship between knee structure and osteoporosis, the dataset enables advanced analyses, such as classification and predictive modeling, and aims to bridge the gap between accessibility and diagnostic accuracy in osteoporosis care.
Furthermore, the dataset includes clinical factors such as age, gender, menopause age, history of diseases such as diabetes and thyroid conditions, fracture history, lifestyle factors, and T-score values derived from Quantitative Ultrasound Systems. In the study, we used only the x-ray version to evaluate the proposed model. 
This dataset \cite{majeed2021knee_dataset_mendeley} features X-ray images specifically targeted at understanding knee osteoporosis. The images are curated to provide a comprehensive view of the condition, making it suitable for various analyses, including classification and predictive modeling. The data is intended for researchers and practitioners aiming to explore the relationship between knee structure and the presence of osteoporosis, offering potential insights into diagnostic practices.

\begin{figure}[htp]
    \centering
    \includegraphics[width=8cm]{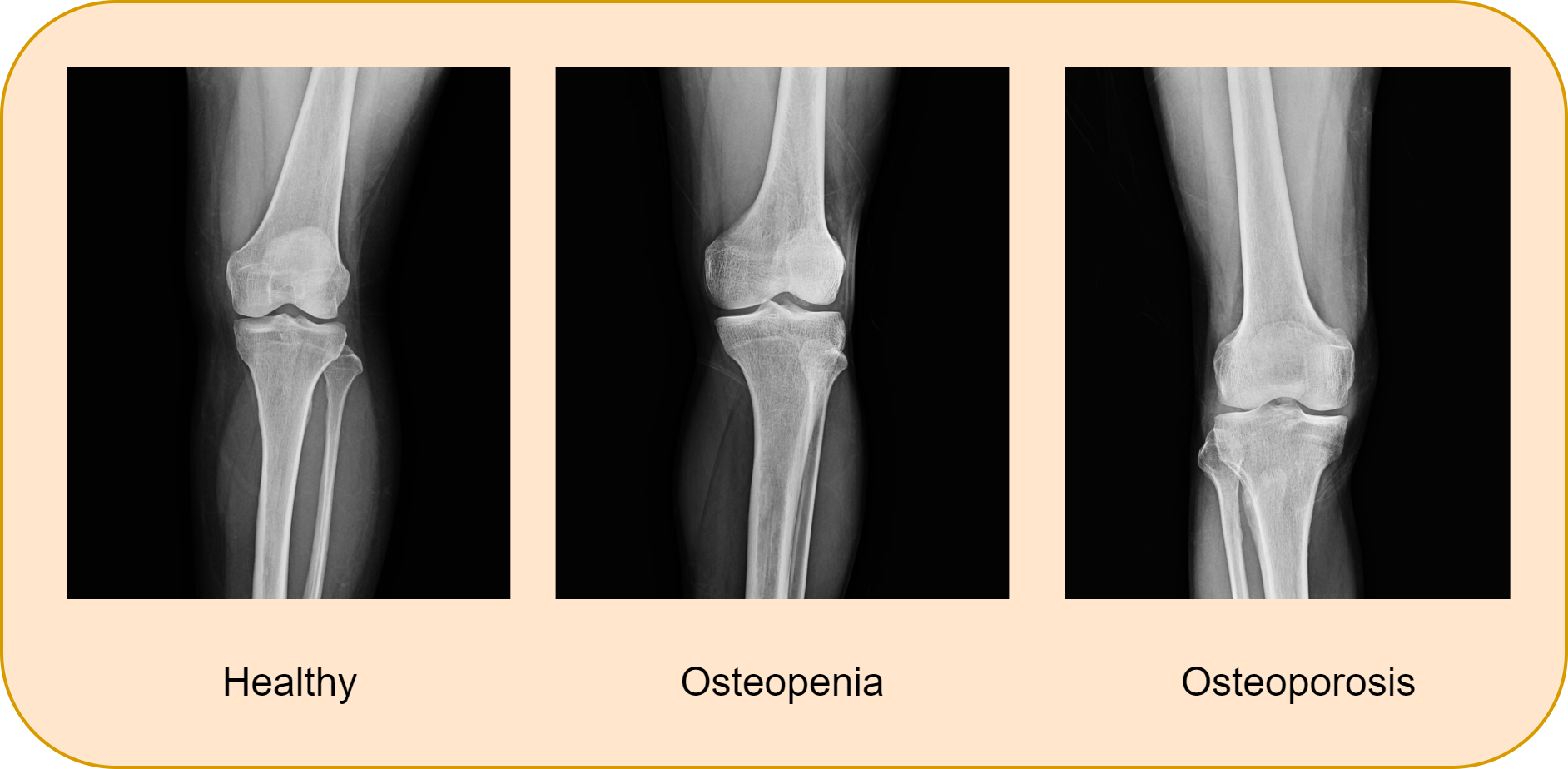}
    \caption{Visual Information of Knee X-rays: Healthy bone and affected by Osteoporosis from KXO-Mendely Multi-Class Dataset and OKX Kaggle Multiclass Dataset}
    \label{fig:Sample Image}
\end{figure}
\subsection{Osteoporosis knee x-ray Kaggle multiclass dataset (OKX Kaggle Multiclass Dataset)}
\label{sec: Kaggle Dataset 2}
%%%https://www.kaggle.com/datasets/mohamedgobara/multi-class-knee-osteoporosis-x-ray-dataset
This dataset \cite{gobara2022knee_dataset_kaggle} comprises a diverse set of knee X-ray images categorized into multiple classes of osteoporosis. It is designed for use in multi-class classification tasks, where the goal is to accurately classify images based on the severity and type of osteoporosis. This dataset is particularly useful for developing robust machine learning algorithms, allowing for improved diagnostic accuracy and enhancing our understanding of knee osteoporosis through detailed analysis of imaging data. In this study, we combined these three different datasets to enhance the robustness and generalizability of the osteoporosis detection model. Combining datasets from various sources helps capture a broader range of variations in knee X-ray images, which can improve the model's ability to generalize across multiple populations and imaging conditions.

\section{Proposed Methodology}
\label{sec: Proposed Methodology}
\begin{figure*}[htp]
    \centering
    \includegraphics[width=15cm]{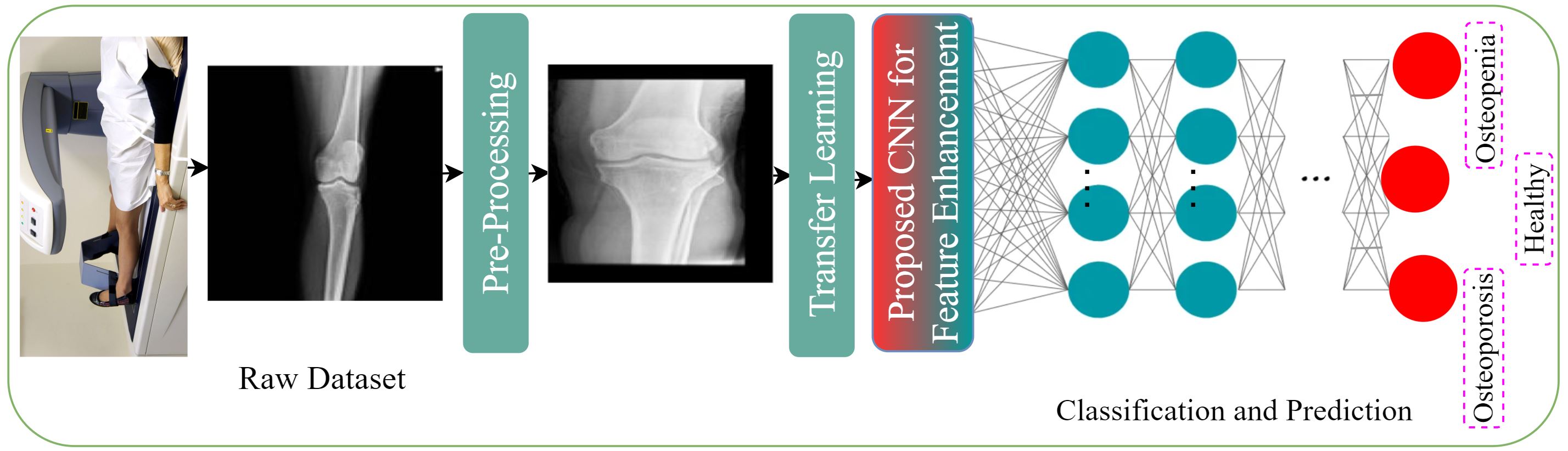}
    \caption{Architecture of the Proposed Methodology}
    \label{fig:Methodology}
\end{figure*}
\begin{figure*}[htp]
    \centering
    \includegraphics[width=16cm]{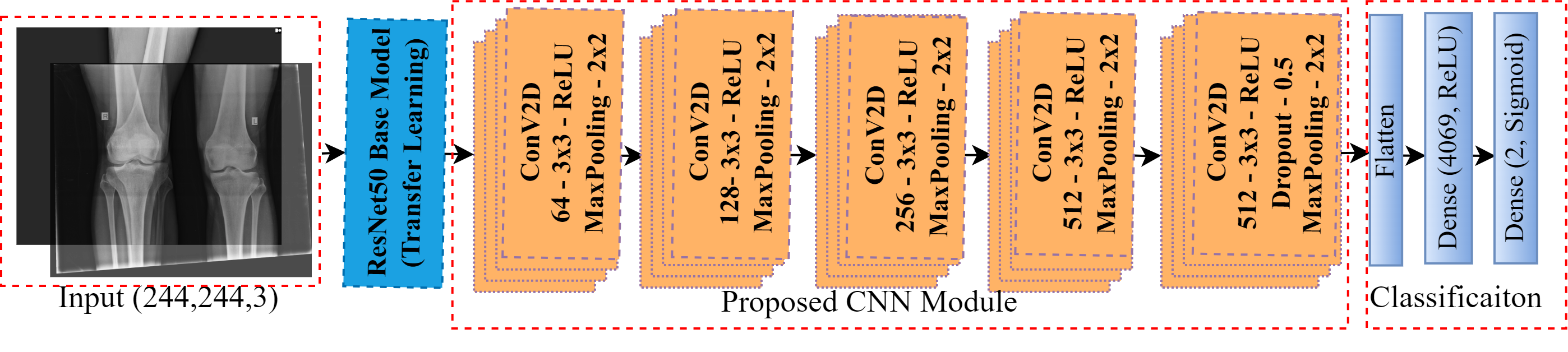}
    \caption{Model Architecture}
    \label{fig:Model Architecture}
\end{figure*}
The figure \ref{fig:Methodology} illustrates the workflow of the proposed computer-aided knee osteoporosis diagnosis with X-ray images using transfer Learning with
Enhanced Features from Stacked Deep
Learning Modules. The proposed methodology for classifying knee X-ray images to diagnose osteoporosis involves a comprehensive approach that includes preprocessing, feature extraction using transfer learning, and feature enhancement using deep learning \cite{cmc.2024.049296_miah_city_name_miah,Bantraffic_miah} blocks, and final classification. Initially, raw knee X-ray images undergo preprocessing, which includes resizing, normalization, and data augmentation to standardize and diversify the input data. Following preprocessing, we employ a pre-trained ResNet50 model for initial feature extraction. By utilizing transfer learning, ResNet50 leverages its pre-learned weights to extract robust and generalizable features from the knee X-ray images, providing a solid foundation for further enhancement. The top fully connected layers of ResNet50 are removed to retain only the convolutional base, which serves as a feature extractor.
To refine and enhance these initial features, we introduce a series of five sequential Conv-RELU-MaxPooling blocks, termed Stacked Feature Enhancement Deep Learning Blocks. Each block consists of a Conv2D layer with 3x3 filters to capture local spatial patterns, a ReLU activation function to introduce non-linearity and a MaxPooling layer with 2x2 filters to downsample the feature maps while retaining important features. These blocks, when stacked sequentially, capture progressively more complex and hierarchical feature representations, crucial for distinguishing between healthy \cite{rahman2023federated} and osteoporotic knees. The enhanced features are then fed into the classification module, which includes a flattened layer to convert 2D feature maps into a 1D vector, followed by dense layers. The first dense layer, with 4096 units and ReLU activation, learns complex feature combinations, while a Dropout layer with a rate of 0.5 prevents overfitting. The final output layer, with two units and a sigmoid activation function, produces probabilities for the classes Osteoporosis and Healthy. Extensive experiments with three individual datasets and a combined dataset demonstrated that our model, integrating transfer learning with stacked feature enhancement blocks, achieved high-performance accuracy, effectively generalizing across diverse input data, reducing overfitting, and increasing diagnostic accuracy in clinical settings. The figure \ref{fig:Model Architecture} depicts the architecture of a customized convolutional neural network (CNN) based on the ResNet50 model.

\subsection{Data Preprocessing}
\label{sec: Data Preprocessing}
%\subsubsection{Data Loading and Preprocessing}
The preprocessing stage prepares the raw knee X-ray images for model training, consisting of three key steps, each of which plays a crucial role in optimizing the input data for the model and enhancing its diagnostic capabilities:

\subsubsection{Resizing}  
All images were resized to a uniform dimension of \(224 \times 224\) pixels to ensure consistency and reduce computational load. This resizing ensures that the input size is standardized for the model, represented as \(224 \times 224 \times 3\), where 224 and 224 represent the height and width, and 3 corresponds to the RGB color channels. The resizing operation can be expressed as:

\begin{equation}
    I'_{\text{resize}} = \text{Resize}(I, 224, 224)
\end{equation}

where \( I \) is the original image and \( I'_{\text{resize}} \) is the resized image. This step reduces the computational burden without significant loss of crucial information, enabling the model to process a consistent input size, which is critical for accurate classification. Resizing also facilitates faster processing, enabling efficient training and inference, which is vital in clinical settings where timely diagnosis is essential.

\subsubsection{Normalization:}  
Pixel values were normalized to the range \([0, 1]\) by dividing each pixel value by 255, which facilitates faster convergence during training. Additionally, images were standardized by subtracting the mean and dividing by the standard deviation, ensuring that the data is zero-centered. The normalization operation is defined as:

\begin{equation}
    I'_{\text{norm}} = \frac{I_{\text{resize}}}{255}
\end{equation}

where \( I_{\text{resize}} \) is the resized image and \( I'_{\text{norm}} \) is the normalized image. Normalization helps in stabilizing and speeding up the training process by making the gradient descent optimization process more efficient. Standardization ensures that the model treats all pixel values equally, preventing any one feature from dominating the learning process, which is crucial for distinguishing subtle differences between healthy and osteoporotic knees.
\subsubsection{Data Augmentation:}  
To enhance the generalization capability of the model and prevent overfitting, various random transformations were applied, including rotations, width and height shifts, shearing, zooming, and flipping. Mathematically, any transformation can be represented as:
\begin{equation}
    I' = T(I)
\end{equation}
where \( I' \) is the augmented image, \( I \) is the original image, and \( T \) is the transformation function. These augmentations expose the model to diverse variations of the input images, allowing it to learn more robust features and improve generalization. This step is particularly important for medical image datasets, as it introduces variability in terms of image orientation, scale, and alignment, simulating real-world scenarios where knee X-ray images may vary due to patient positioning or image acquisition conditions. By artificially expanding the dataset, data augmentation helps the model generalize better across unseen data and improves its ability to make accurate diagnoses, particularly when working with limited or imbalanced datasets. The augmented images are then used to train the model, improving its ability to generalize across different input variations. Each of these preprocessing steps—resizing, normalization, and augmentation—contributes to enhancing the performance and robustness of the model, making it more reliable for accurate knee osteoporosis diagnosis from X-ray images.

\subsection{Transfer Learning Based Feature Extraction }
\label{sec: Feature Extraction}In this study, we utilize the ResNet50 architecture for feature extraction from knee X-ray images to detect osteoporosis \cite{siddiqua2024transfer}. The process begins with the Input Layer, which accepts images with dimensions \( 224 \times 224 \times 3 \), where \( 224 \) and \( 224 \) represent the height and width of the image, and \( 3 \) corresponds to the RGB color channels. These resized images are then passed through the ResNet50 Base Model, which has been pre-trained on the ImageNet dataset. By leveraging pre-learned weights, the ResNet50 model provides powerful feature representations that help improve the efficiency of training and enhance the model's performance, especially when fine-tuning for osteoporosis detection. Technically, this model consisted with convolutional layers, maxpooling and residual block which we described below. 

\subsubsection{Convolutional Layers}
Each convolutional layer in ResNet50 uses a \( 3 \times 3 \) filter to convolve with the input image. The convolution operation at the \( k \)-th feature map location is mathematically represented as:

\begin{equation}
    I_{\text{conv}}(i,j,k) = \sum_{m=-1}^{1} \sum_{n=-1}^{1} I_{\text{input}}(i+m, j+n, k) \cdot W(m,n)
\end{equation}

where:
- \( I_{\text{conv}}(i,j,k) \) is the output of the convolutional operation at position \( (i, j) \) for the \( k \)-th feature map.
- \( I_{\text{input}}(i+m, j+n, k) \) is the input image at position \( (i+m, j+n) \), for the \( k \)-th channel.
- \( W(m,n) \) represents the weights of the filter.

After convolution, the output is passed through the ReLU activation function:

\begin{equation}
    I_{\text{relu}}(i,j,k) = \max(0, I_{\text{conv}}(i,j,k))
\end{equation}

where \( I_{\text{relu}}(i,j,k) \) is the output after applying the ReLU activation.

\subsubsection{MaxPooling}
Each convolutional block is followed by a MaxPooling layer with a \( 2 \times 2 \) filter. The pooling operation can be represented as:

\begin{equation}
    I_{\text{pool}}(i,j,k) = \max_{m,n} (I_{\text{input}}(i+m, j+n, k))
\end{equation}

where:
- \( I_{\text{pool}}(i,j,k) \) is the pooled output at position \( (i,j) \) for the \( k \)-th feature map.
- The pooling operation selects the maximum value from a \( 2 \times 2 \) neighborhood of the input feature map.

MaxPooling helps in reducing spatial dimensions and retaining the most important features, which are essential for capturing the relevant aspects of knee X-ray images.

\subsubsection{ Residual Blocks}
The ResNet50 architecture utilizes residual blocks to address the vanishing gradient problem. Each residual block consists of two or more convolutional layers, and a shortcut connection bypasses the convolutional layers. The output of a residual block can be written as:

\begin{equation}
    \text{Output}_{\text{res}} = F(x) + x
\end{equation}

where:
- \( F(x) \) represents the feature map produced by the convolutional layers in the residual block.
- \( x \) is the input to the residual block, and the addition operation represents the shortcut connection.

The residual connection allows for easier gradient flow, enabling the network to learn deeper representations without suffering from vanishing gradients.

\subsubsection{Final Feature Output of ResNet50}
After the application of the convolutional layers, ReLU activation, MaxPooling, and residual connections, the final output of the ResNet50 base model, which serves as the feature representation for the input image, is denoted as \( F_{\text{ResNet50}} \):
\begin{equation}
    F_{\text{ResNet50}} = \mathcal{P} \left( \sum_{l=1}^{L} \left( \text{ResBlock}_l \left( \text{Conv}_l ( I_{\text{input}} ) \right) + \text{Shortcut}_l \right) \right)
\end{equation}Where:
$F_{\text{ResNet50}}$ is the final output feature map from ResNet50. $I_{\text{input}}$ is the input image with size $224 \times 224 \times 3$. $L$ represents the total number of residual blocks in ResNet50.
$\text{Conv}_l$ refers to the convolutional layers in the $l$-th residual block.
$\text{ResBlock}_l$ represents the operations within the $l$-th residual block, including convolution, batch normalization, and ReLU activation.
$\text{Shortcut}_l$ is the residual shortcut connection that bypasses the convolutional operations in each block.
$P$ represents the pooling operation (e.g., max pooling) applied to the final feature map.
This equation captures the entirety of the ResNet50 process, from the input image through the stacked residual blocks, with their corresponding convolutions, activations, and pooling layers, up to the final feature map output. 
Convolution Layers perform convolutions on the input image, progressively extracting higher-level features such as edges, textures, and shapes. Residual Blocks Each block consists of multiple convolutional layers and a shortcut connection that helps the model avoid the vanishing gradient problem. Shortcut Connections adds the input of each residual block directly to the output, facilitating the training of deeper models by mitigating the vanishing gradient problem. Pooling layers reduce spatial dimensions and preserve important features, allowing the network to focus on high-level characteristics while reducing computational complexity. The output feature map \( F_{\text{ResNet50}} \) from ResNet50 is then fed into the subsequent Stack feature enhancement module,  as detailed in the next section of the methodology.

\subsection{ Stack Feature Enhancement  Module}
\label{sec: Proposed CNN Module}
The Stack Feature Enhancement Module consists of five sequential blocks, each comprising a convolutional layer, a ReLU activation function, and a MaxPooling layer. Each block progressively refines the feature map, capturing hierarchical patterns and enabling efficient learning for osteoporosis classification. Below, we describe each block along with its corresponding mathematical equations.

\subsubsection{Feature Enhancement Block 1}
This block take the output of the transfer learning as input and enhance them as below: 

\begin{itemize}
    \item \textbf{Convolutional Layer:} The input feature map \( F_{\text{ResNet50}} \) from the pre-trained ResNet50 model is processed using 64 filters with a kernel size of \( 3 \times 3 \):
    \begin{equation}
        F_{\text{conv1}} = \text{ReLU}(\text{Conv}(F_{\text{ResNet50}}, W_1) + b_1),
    \end{equation}
    where \( W_1 \) and \( b_1 \) are the weights and biases of the first convolutional layer.
    \item \textbf{Pooling Layer:} MaxPooling is applied to down-sample the feature map:
    \begin{equation}
        F_{\text{pool1}} = \text{MaxPool}(F_{\text{conv1}}, k=2, s=2),
    \end{equation}
    where \( k=2 \) is the pooling kernel size, and \( s=2 \) is the stride.
\end{itemize}
This block ensures the extraction of low-level spatial features, such as edges and textures, from the X-ray images, forming the foundation for capturing more complex patterns in later layers.
\subsubsection{Feature Enhancement Block 2}
The output from the first block, \( F_{\text{pool1}} \), is processed with 128 filters:
    \begin{equation}
        F_{\text{conv2}} = \text{ReLU}(\text{Conv}(F_{\text{pool1}}, W_2) + b_2),
    \end{equation}
    where \( W_2 \) and \( b_2 \) are the weights and biases of the second convolutional layer.
 MaxPooling is applied to down-sample the feature map:
    \begin{equation}
        F_{\text{pool2}} = \text{MaxPool}(F_{\text{conv2}}, k=2, s=2).
    \end{equation}
This block extracts mid-level features, such as patterns related to bone density and localized irregularities, enabling the network to learn more detailed characteristics.

\subsubsection{Feature Enhancement Block 3}
The second block's output, \( F_{\text{pool2}} \), is processed with 256 filters:
    \begin{equation}
        F_{\text{conv3}} = \text{ReLU}(\text{Conv}(F_{\text{pool2}}, W_3) + b_3),
    \end{equation}
    where \( W_3 \) and \( b_3 \) are the weights and biases of the third convolutional layer.
MaxPooling is applied:
    \begin{equation}
        F_{\text{pool3}} = \text{MaxPool}(F_{\text{conv3}}, k=2, s=2).
    \end{equation}

This block captures finer details such as small structural irregularities and early signs of osteoporosis progression, enhancing the model's sensitivity to critical markers.

\subsubsection{Feature Enhancement Block 4}
The third block's output, \( F_{\text{pool3}} \), is processed with 512 filters:
    \begin{equation}
        F_{\text{conv4}} = \text{ReLU}(\text{Conv}(F_{\text{pool3}}, W_4) + b_4),
    \end{equation}
    where \( W_4 \) and \( b_4 \) are the weights and biases of the fourth convolutional layer.
  Down-sampling is applied:
    \begin{equation}
        F_{\text{pool4}} = \text{MaxPool}(F_{\text{conv4}}, k=2, s=2).
    \end{equation}
This block extracts high-level, domain-specific features, such as joint deformities and severe thinning, ensuring the model focuses on markers critical for distinguishing between healthy and osteoporotic conditions.

\subsubsection{Representation of the Stack Module Features}
The output of the fourth block, \( F_{\text{pool4}} \), undergoes a convolution operation with 512 filters:
    \begin{equation}
        F_{\text{conv5}} = \text{ReLU}(\text{Conv}(F_{\text{pool4}}, W_5) + b_5),
    \end{equation}
    where \( W_5 \) and \( b_5 \) are the weights and biases of the fifth convolutional layer.
MaxPooling reduces the spatial dimensions of the feature map:
    \begin{equation}
        F_{\text{pool5}} = \text{MaxPool}(F_{\text{conv5}}, k=2, s=2),
    \end{equation}
    where \( k=2 \) and \( s=2 \) are the kernel size and stride of the pooling layer.
To prevent overfitting, a Dropout layer with a rate of 0.5 is applied:
    \begin{equation}
        F_{\text{final}} = \text{Dropout}(F_{\text{pool5}}, p=0.5),
    \end{equation}
    where \( p=0.5 \) is the probability of randomly disabling 50\% of the neurons during each training step.
The initial feature map \( F_{\text{ResNet50}} \) is extracted from the pre-trained ResNet50 model and processed sequentially through five enhancement blocks, each designed to progressively refine the feature map. Starting with low-level spatial details, the convolutional layers extract hierarchical features, ReLU activation introduces non-linearity for complex pattern recognition, and pooling layers reduce dimensions while preserving salient information. The integration of a Dropout layer in the fifth block mitigates overfitting, ensuring robust generalization.
The final feature map \( F_{\text{final}} \), enriched with domain-specific patterns, is then fed into the classification module for accurate diagnosis.

\subsection{Classification Module}
% \label{sec: Classification Module}
The classification module processes the final feature map \( F_{\text{final}} \), generated by the Stack Feature Enhancement Module, to produce class predictions. It consists of a Flatten layer, two fully connected Dense layers, and an output layer, designed specifically to enhance the diagnosis of X-ray-based diseases such as knee osteoporosis. The detailed description is as follows:

\subsubsection{Flatten Layer} The Flatten layer converts the 2D feature map \( F_{\text{final}} \) into a 1D vector \( \mathbf{f}_{\text{final}} \), preparing it for input into the fully connected layers:
    \begin{equation}
        \mathbf{f}_{\text{final}} = \text{Flatten}(F_{\text{final}}).
    \end{equation}
    This Flattening ensures that spatial and hierarchical features extracted from X-ray images are preserved and formatted correctly for further processing, facilitating the accurate detection of disease-relevant patterns.

\subsubsection{First Dense Layer} The first Dense layer consists of 4096 units with a ReLU activation function, enabling the model to learn complex feature representations:
    \begin{equation}
        \mathbf{h}_1 = \text{ReLU}(\mathbf{f}_{\text{final}} \cdot W_1 + b_1),
    \end{equation}
    where \( W_1 \) and \( b_1 \) are the weights and biases of the first Dense layer. \\
    This layer combines features extracted from the enhancement blocks, capturing nuanced patterns like bone thinning or irregular joint spaces, which are critical for diagnosing osteoporosis in X-ray images.

\subsubsection{Second Dense Layer} The second Dense layer further refines the feature representations with another 4096 units and ReLU activation:
    \begin{equation}
        \mathbf{h}_2 = \text{ReLU}(\mathbf{h}_1 \cdot W_2 + b_2),
    \end{equation}
    where \( W_2 \) and \( b_2 \) are the weights and biases of the second Dense layer. \\
   This layer enhances the model’s ability to distinguish subtle differences in bone structures, such as variations in density and texture, crucial for accurate classification in X-ray-based systems.

The final Dense layer consists of 2 units and uses a Sigmoid activation function to output probabilities for the two classes (Healthy and Osteoporosis):
    \begin{equation}
        \mathbf{p} = \text{Sigmoid}(\mathbf{h}_2 \cdot W_3 + b_3),
    \end{equation}
    where \( W_3 \) and \( b_3 \) are the weights and biases of the output layer, and \( \mathbf{p} = [p_1, p_2] \) represents the probabilities for the two classes. \\
   The Sigmoid function provides normalized probabilities, enabling the model to make confident binary predictions while capturing the subtle indicators of disease.

\subsubsection{Final Prediction}
The class probabilities are computed as:
\begin{equation}
    p_i = \frac{1}{1 + e^{-(\mathbf{h}_2 \cdot W_3 + b_3)}} \quad \text{for } i = 1, 2.
\end{equation}

The final predicted class \( \hat{y} \) is determined as:
\begin{equation}
    \hat{y} = \arg\max(\mathbf{p}),
\end{equation}
where \( \hat{y} \in \{ \text{Healthy}, \text{Osteoporosis} \} \).

The classification module integrates hierarchical feature representations from the enhancement blocks with dense layers, enabling precise diagnosis of X-ray-based diseases. The Flatten layer preserves spatial and hierarchical features, the Dense layers enhance feature representation and abstraction, and the Sigmoid output provides robust class probabilities. This design ensures high sensitivity and specificity in distinguishing between Healthy and Osteoporotic knee conditions.

% The Classification Module is responsible for making the final prediction based on the features extracted and processed by the earlier layers of the network. In this architecture, after the features are extracted through the convolutional and fully connected layers, they are passed to the classification block. This module consists of a fully connected layer with 4096 units and ReLU activation, which processes the learned features and captures complex patterns and relationships within the data. The output of this dense layer is then fed into the output layer, which has 2 units corresponding to the two possible classes: "Healthy" and "Osteoporotic." A sigmoid activation function is used in the output layer to generate probabilities for each class. The model assigns the class with the higher probability as the final prediction, effectively classifying the input knee X-ray image as either "Healthy" or "Osteoporotic." This classification module ensures the model's ability to make accurate and reliable predictions based on the learned features.

\subsection{Model Training and Hyperparameter Tuning}
\label{sec: Model Training and Hyperparameter Tuning}
The training of the proposed model was conducted with a series of hyperparameter tuning to optimize its performance. Key hyperparameters such as the learning rate, batch size, number of epochs, and dropout rate were carefully adjusted to achieve the best results. We employed a grid search approach to explore various combinations of learning rates and batch sizes, aiming to find the optimal configuration. The model was trained for 50 epochs with a batch size of 32, and the Adam optimizer was used with an initial learning rate of 0.001. Dropout was applied at a rate of 0.5 to mitigate overfitting. The impact of these hyperparameters was evaluated using cross-validation, ensuring that the model generalized well to unseen data. The final tuned model achieved improved accuracy and robustness, as discussed in the results section.

In summary, transfer learning in this study capitalizes on the pre-trained EfficientNetB0 model to leverage learned features from a large and diverse dataset. By combining this pre-trained model with custom layers, we efficiently adapt the network to our specific problem, improving both training efficiency and model performance. This approach allows us to achieve high accuracy with reduced training time and computational resources compared to training a model from scratch.

\begin{comment}
    Data augmentation was employed to artificially expand the training dataset and enhance the model's ability to generalize. Several transformations were applied to achieve this. Rotation involved randomly rotating images by up to 10 degrees, introducing variations in orientation. Width and Height Shift involved random shifts of images by up to 10% of their width and height, altering their position within the frame. Shear transformations were applied randomly up to 10 degrees, skewing the images to simulate different perspectives. Zoom adjustments allowed for random zooming by up to 10%, varying the scale of the images. Finally, Horizontal Flip was used to randomly flip images horizontally, creating mirrored versions of the images. These augmentations help the model to learn invariant features by exposing it to a variety of transformations of the input images.
\end{comment}

\section{Experimental Evaluation}
\label{sec: Experimental Evaluation}
We evaluate the proposed enhanced model using publicly available datasets to achieve promising results in terms of accuracy. Datasets are consist of labeled images of healthy, osteopenia and osteoporotic knees. The dataset was preprocessed by resizing all images to 224×224×3 and normalizing pixel values to the range [0,1]. We divided the data into training, validation and test sets, with 60\%, 20\%, and 20\% of the data used for each, respectively.
%The dataset was divided into three distinct subsets to facilitate effective model training and evaluation. The Training Set comprised 60\% of the total dataset, serving as the primary source of data for training the model. This substantial portion ensured that the model had ample examples to learn from, allowing it to capture and generalize patterns effectively. The Validation Set constituted 20\% of the total dataset and was used to tune model hyperparameters and evaluate the model's performance during training. This subset helped in monitoring and improving the model's performance iteratively. Lastly, the Test Set also made up 20\% of the total dataset and was reserved for the final evaluation of the model's performance. This set provided an unbiased assessment of the model's accuracy and generalization capabilities on new, unseen data. 

\subsection{Environmental Setting for Experiments}
\label{Environment Setup}
For this experiment, we used Google Colab as the coding interpreter and Python 3.0 as the programming language. We utilized the integrated 16 GB of RAM and 12 GB of GPU throughout the training process. The environment configuration is given in Table \ref{sec: Environment Configuration}. 

\begin{table}[]
\caption{Environment Configuration}
\label{sec: Environment Configuration}
\begin{tabular}{llll}
\hline
\textbf{Component} &  &  & \textbf{Specifications}                           \\ \hline
Execution Platform &  &  & Google Colab                                     \\
Programming Language &  &  & Python 3                                         \\
Memory (RAM)       &  &  & 16 GB                                            \\
Graphics Processing Unit (GPU) &  &  & TPU 2.0 12 GB \\
Key Libraries      &  &  & Pytorch, Tensorflow, Cv2            \\
Storage (Disk Space) &  &  & 5 GB utilized                                    \\ \hline
\end{tabular}
\end{table}
\subsection{Model Evaluation Metrics}
\label{sec: Model Evaluation Metrics} 
The model was trained and evaluated on a comprehensive test set comprising diverse datasets to enhance osteoporosis detection across varying imaging conditions. Key metrics included accuracy, precision, recall, and F1 score  \cite{hicks2022evaluation}. \textbf{Accuracy} represents the proportion of correctly classified images, while \textbf{precision} is the ratio of true positives to the total predicted positives:
\[
Precision = \frac{TP}{TP + FP}
\]
Recall measures the ratio of true positives to the total actual positives:
\[
Recall = \frac{TP}{TP + FN}
\]
The F1 score, as the harmonic mean of precision and recall, is calculated as:
\[
F1 Score = 2 \times \frac{\text{Precision} \times \text{Recall}}{\text{Precision} + \text{Recall}}
\]
A classification report detailing precision, recall, and F1 score was generated for each class to provide additional performance insights. A confusion matrix visualized true versus predicted classifications, summarizing true positives, true negatives, false positives, and false negatives for an intuitive understanding of model strengths and areas for improvement. The training and validation accuracy and loss curves were plotted to assess the model's learning progress over the epochs. These curves visualize the recorded accuracy and loss values for each epoch, providing insights into how well the model fits the training data and generalizes to the validation set.

\subsection{Performance Result with Binary Classes}
\label{sec: Experimental Results of the Proposed Model}
%OKX Kaggle Binary Dataset
%KXO-Mendely Binary Dataset
%KXO-Mendely Multi-Class Dataset
%OKX Kaggle Binary Dataset
%OKX Kaggle Multiclass Dataset
Performance evaluation of the proposed model for binary classification tasks demonstrates its robust and reliable results across multiple datasets. We evaluated the performance of our proposed model for three datasets in terms of overall metrics shown in Table \ref{tab:binary_classification_results}. To evaluate the performance of the model, we used accuracy, precision, recall, and the F1 score. These metrics provide a comprehensive view of the model's ability to classify the images into the two categories correctly. 
The proposed model demonstrates strong performance in binary classification, as shown in Table \ref{tab:binary_classification_results}. In the OKX Kaggle Binary Dataset, the model achieved a precision of 97.20\%, a recall of 97.30\%, an F1-score of 97.32\%, and an accuracy of 97.32\%. For the KXO-Mendely Multiclass Dataset, the model achieved a precision of 98.40\%, a recall of 98.40\%, an F1-score of 98.40\%, and an accuracy of 98.24\%. In the OKX Kaggle Multiclass Dataset, the metrics further improved, with precision at 97.10\%, recall at 97.20\%, F1-score at 96.40\%, and accuracy at 97.27\%. For the Combined Dataset, the model excelled, achieving its highest performance metrics: 97.70\% precision, 97.70\% recall, 97.70\% F1-score, and an accuracy of 98.00\%. These results confirm the reliability and effectiveness of the model in handling binary classification tasks in diverse datasets. 
\begin{table*}[!htp] 
\centering
\caption{Performance metrics of the proposed model for binary-class classification: accuracy, precision, F1-score, and recall.} 
\label{tab:binary_classification_results} 
\begin{tabular}{|l|l|l|l|l|} 
\hline  
 & Precision & Recall & F1-score & \begin{tabular}[c]{@{}l@{}} Performance \\ Accuracy\end{tabular}  \\ \hline 
OKX Kaggle Binary Classification & 97.20  &  97.30 & 97.32 & 97.32 \\ \hline
KXO-Mendely Multi-Class Dataset  & 98.40 &  98.40  & 98.40  &  98.24 \\ \hline
OKX Kaggle Multiclass Dataset & 97.10 & 97.20   & 96.40  & 97.27  \\ \hline
Combined Dataset & 97.70 &  97.70  & 97.70  &  98.00 \\ \hline
\end{tabular} 
\end{table*}

\begin{figure*}[!t]
     \centering
     \begin{subfigure}[t]{0.35\textwidth}
     \centering
     \includegraphics[width=7.5cm]{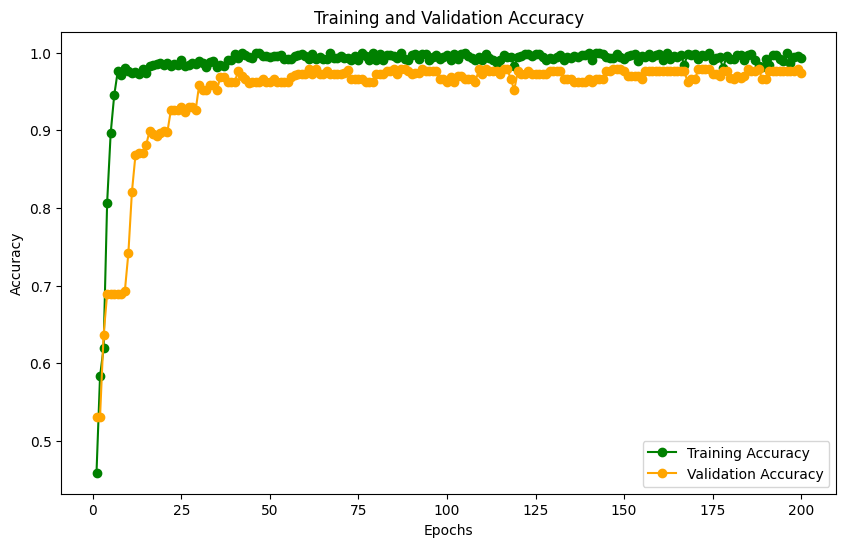}
     \caption{Training and Validation Accuracy curve}
    \label{fig:Training_Accuracy_Dataset-1}
     \end{subfigure}
    \hfill
    \begin{subfigure}[t]{0.56\textwidth}
         \centering
         \includegraphics[width=9cm]{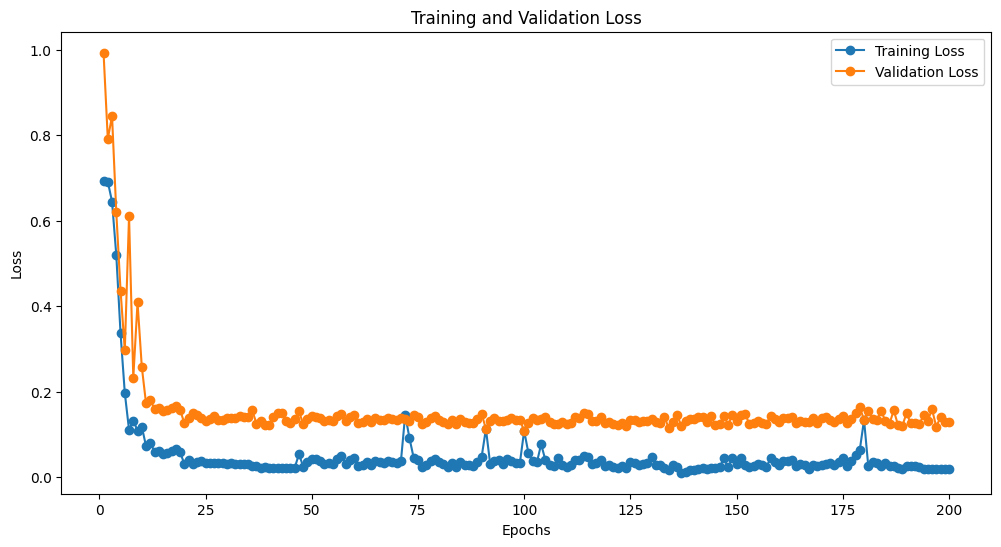}
         \caption{Training and Validation Loss curve}
         \label{fig:Training_loss_Dataset-1}
     \end{subfigure}
     \caption{(a) Training and Validation Accuracy curve, and (b) Training and Validation Loss curve for OKX Kaggle Binary Classification}
     \label{fig:OKX Kaggle Binary Classification}
\end{figure*}

\begin{figure*}[!t]
     \centering
     \begin{subfigure}[t]{0.49\textwidth}
     \centering
     \includegraphics[width=7cm]{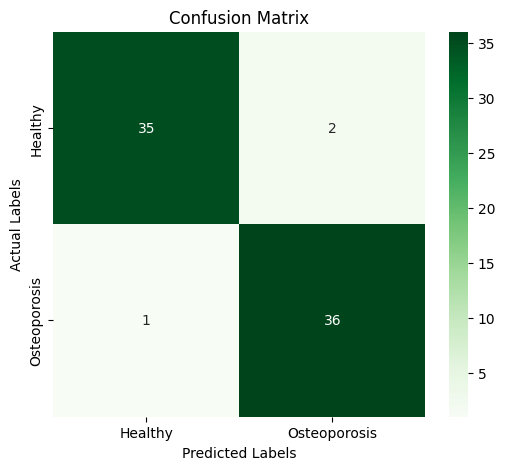}
     \caption{OKX-Kaggle Binary Dataset}
    \label{fig:Confusion_matrix_Dataset-1}
     \end{subfigure}
    \hfill
    \begin{subfigure}[t]{0.49\textwidth}
         \centering
         \includegraphics[width=7cm]{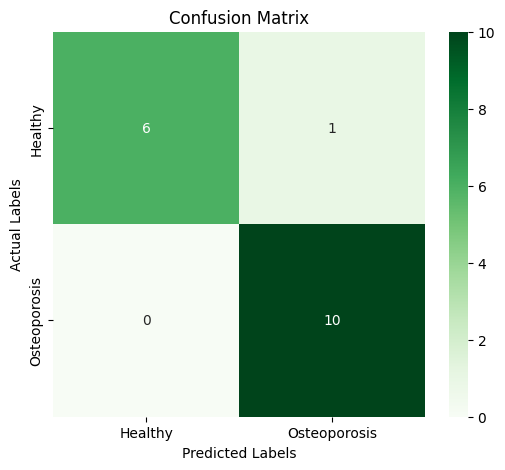}
         \caption{KXO-Mendely Binary-class Dataset}
         \label{fig:Confusion_matrix_Dataset-2}
     \end{subfigure}
     \hfill
     \begin{subfigure}[t]{0.49\textwidth}
         \centering
         \includegraphics[width=7cm]{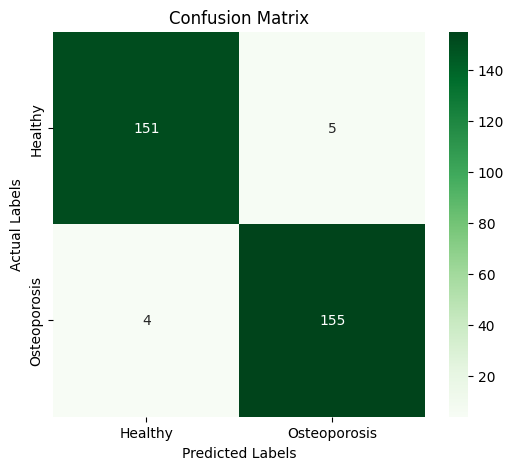}
         \caption{OKX-Kaggle binary-class Dataset}
         \label{fig:Confusion_matrix_Dataset-3}
     \end{subfigure}
     \hfill
     \begin{subfigure}[t]{0.49\textwidth}
         \centering
         \includegraphics[width=7cm]{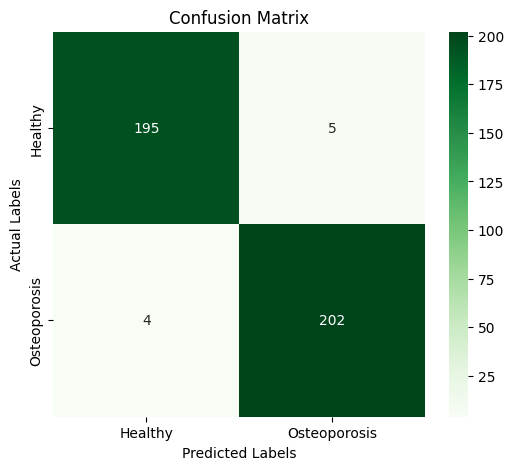}
         \caption{Combined Binary Dataset}
         \label{fig:Combined_Confusion matrix}  
     \end{subfigure}
     \caption{ Confusion matrix for Binary Classification (a) OKX Kaggle Binary Dataset, (b) KXO-Mendely Multi-class Dataset, (c) OKX-Kaggle Multi-class Dataset, and (d) Combined Dataset. }
     \label{fig:Confusion matrix for Binary Classification}
\end{figure*}

Figures \ref{fig:Confusion_matrix_Dataset-1}, \ref{fig:Training_Accuracy_Dataset-1}, \ref{fig:Training_loss_Dataset-1} demonstrated the confusion matrix, the accuracy curve, and the loss curve of the OKX Kaggle binary dataset for the proposed model. The confusion matrix indicates that the model effectively classifies 35 samples as healthy and 36 as osteoporosis, demonstrating strong precision and recall, for the Healthy and Osteoporosis classes. The accuracy curve shows that the training accuracy is nearly 100\% and the validation accuracy stabilizes around 98\%, suggesting effective learning and a strong generalization to unseen data, with minimal overfitting. The loss plot reveals a steady decrease in training loss converging near 0, while validation loss stabilizes just above 0, confirming the model's reliability and efficiency in multiclass classification of knee X-ray images. \\

%%%%%%%%%%%%% Confusion Matrix for binary classes %%%%%%%%%%%%%%%%%
%%%%%%%%KXO-Mendely Binary Dataset%%%%%%%%%%
Figure \ref{fig:Confusion_matrix_Dataset-2}, \ref{fig:Training_Accuracy_Dataset-2}, \ref{fig:Training_loss_Dataset-2} demonstrated the confusion matrix, accuracy curve and loss curve of the  OKX Kaggle Binary dataset for the proposed model. These figures demonstrate the strong performance of the model. The confusion matrix highlights high classification accuracy, with 6 out of 7 healthy samples and all 10 osteoporosis samples correctly classified, showcasing excellent sensitivity and specificity. The training and validation accuracy curves show rapid improvement, stabilizing near 100\% and slightly lower for validation, indicating excellent generalization. The loss curves reveal a steady decrease, with training loss nearing zero and validation loss stabilizing with minimal fluctuations, reflecting effective learning and minor overfitting.  \\
%%%%%% OKX Kaggle Binary Dataset %%%%%%%
Figure \ref{fig:Confusion_matrix_Dataset-3}, \ref{fig:Training_Accuracy_Dataset-3}, \ref{fig:Training_loss_Dataset-3} demonstrated the confusion matrix, accuracy curve and loss curve of the  OKX Kaggle Binary dataset for the proposed model. The model's performance indicates a robust classification capability, correctly identifying 151 samples as Healthy and 155 as Osteoporosis, demonstrating high precision and recall for both categories. The training accuracy nearly reaches 100\%, while the validation accuracy stabilizes around 98\%, showcasing effective learning and strong generalization with minimal overfitting. Furthermore, the training loss decreases steadily to almost zero, and the validation loss stabilizes just above zero, confirming the model's efficiency and reliability in the multiclass classification of knee X-ray images. These metrics confirm the model's ability to differentiate between two classes effectively with high predictive accuracy and minimal errors.
\\
%%%%%%% Combined Binary dataset Dataset %%%%%%%
 Figures \ref{fig:Combined_Confusion matrix}, \ref{fig:Combined_Training and Validation Accuracy}, \ref{fig:Combined_Training and Validation Loss} demonstrated the confusion matrix, accuracy curve, and loss curve of the Combined Multi-Class dataset for the proposed model. The confusion matrix in figure\ref{fig:Combined_Confusion matrix}  highlights the model's strong performance across three classes: Healthy, Osteopenia, and Osteoporosis, with high diagonal dominance indicating excellent classification accuracy. Healthy and Osteoporosis classes show outstanding results with 195 and 202 correct predictions, respectively, and minimal misclassifications. It still achieves commendable accuracy. In general, the low off-diagonal values confirm the high sensitivity and specificity of the model, demonstrating reliable performance across all classes. The training and validation accuracy curve \ref{fig:Combined_Training and Validation Accuracy} shows that training accuracy quickly reaches about 100\% by the 25-epoch mark, while validation accuracy stabilizes around 99\%. As training continues, both metrics level off, indicating that the model has converged and optimized performance, with minimal gains beyond 25 to 50 epochs. Overall, the results highlight the robustness of the model in classifying knee X-ray images. The figure \ref{fig:Combined_Training and Validation Loss} illustrates the loss trends for training and validation datasets over 200 epochs, highlighting the model's learning efficiency. The training loss decreases steadily to near-zero, demonstrating the model's excellent fit to the training data. The validation loss stabilizes at a low value, indicating strong generalization and minimal overfitting, ensuring reliable performance on unseen data.

%%%%%%%%%%%%% Confusion Matrix for Multi classes %%%%%%%%%%%%%%%%%
\begin{figure*}[!t]
     \centering
     \begin{subfigure}[t]{0.25\textwidth}
     \centering
     \includegraphics[width=5.7cm]{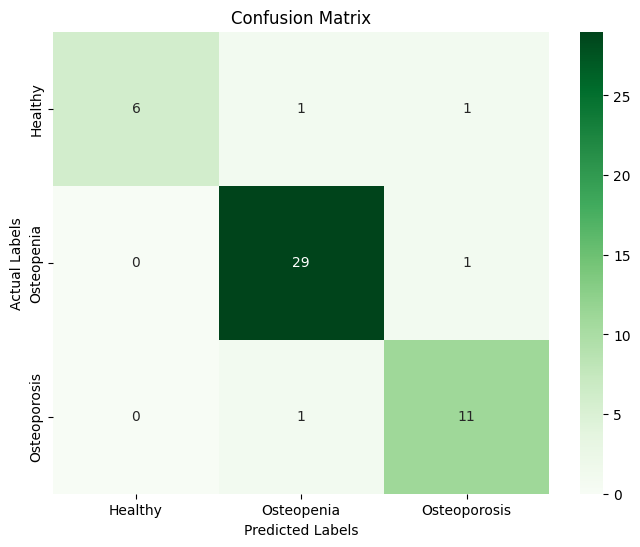}
     \caption{KXO-Mendely Multi-class Dataset}
    \label{fig:mendely_multiclasss_Confusion_matrix_Dataset}
     \end{subfigure}
    \hfill
    \begin{subfigure}[t]{0.25\textwidth}
         \centering
         \includegraphics[width=5.7cm]{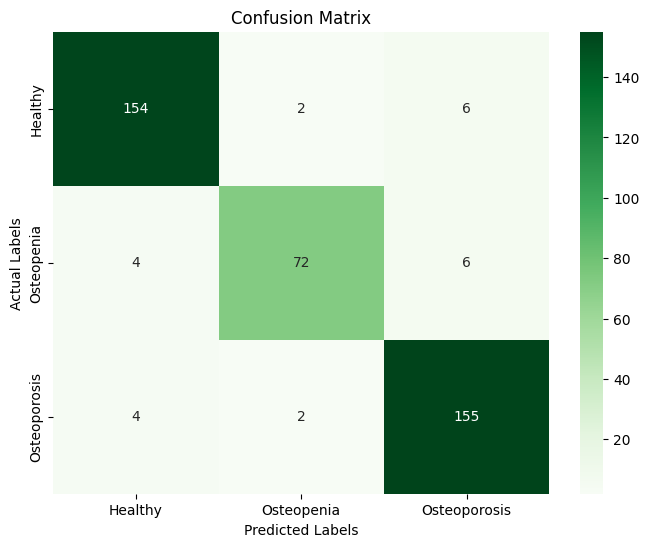}
         \caption{OKX-Kaggle Multi-class Dataset}
         \label{fig: Kaggle Multiclass_Confusion_matrix_Dataset}
     \end{subfigure} 
     \hfill
     \begin{subfigure}[t]{0.25\textwidth}
         \centering
         \includegraphics[width=5.7cm]{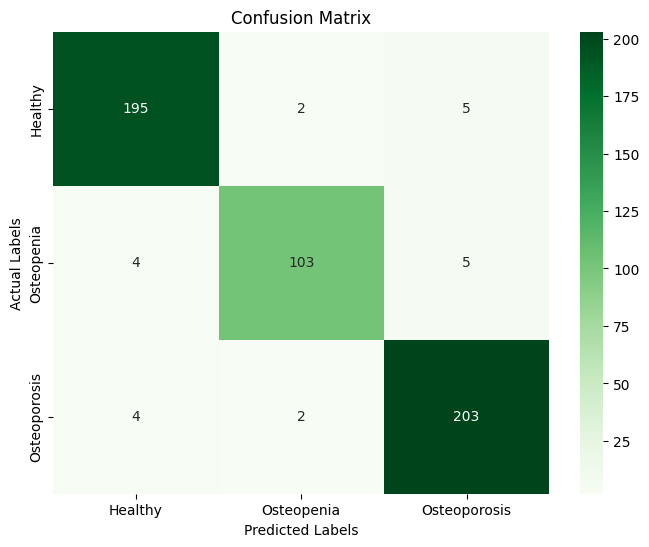}
         \caption{Combined Dataset}
         \label{fig:combined_multiclass_Confusion_matrix_Dataset}
     \end{subfigure}
     \caption{ Confusion matrix for Multi-class Classification (a) KXO-Mendely Multi-class Dataset, (b) OKX-Kaggle Multi-class Dataset, and (c) Combined Dataset. }
     \label{fig:Confusion matrix for Binary Classification}
\end{figure*}

\subsection{Performance Result with Multi-Classes}
\label{sec:Performance Result with Multi-Classes }
The proposed model demonstrates strong performance in multi-class classification, as shown in Table \ref{tab:multi_class_accuracy}. For the KXO-Mendeley Multi-Class Dataset, the model achieved a precision of 97.80\%, recall of 92.70\%, an F1-score of 99.60\%, and accuracy of 98.24\%. On the OKX Kaggle Multiclass Dataset, the metrics further improved, with precision at 93.00\%, recall at 94.20\%, F1-score at 93.60\%, and accuracy at 97.27\%. For the Combined Dataset, the model excelled, achieving its highest performance metrics: 95.20\% precision, 95.90\% recall, 95.50\% F1-score, and an accuracy of 98.40\%. These results confirm the model’s reliability and effectiveness in handling multi-class classification tasks across diverse datasets. These findings highlight the model's robustness and its potential application in clinical settings, where accurate classification of bone health can significantly aid in early diagnosis and treatment planning. 

\begin{table*}[!htp] 
\centering
\caption{Performance matrix of the proposed model for the multi-class classification: accuracy, precision, F1-score, and recall.} 
\label{tab:multi_class_accuracy} 
\begin{tabular}{|l|l|l|l|l|} 
\hline  
 & Precision & Recall & F1-score & \begin{tabular}[c]{@{}l@{}} Performance \\ Accuracy\end{tabular}  \\ \hline 
KXO-Mendely Multi-Class Dataset  & 97.80 & 92.70  & 99.60  & 98.24 \\ \hline
OKX Kaggle Multiclass Dataset & 93.00 & 94.20   & 93.60  & 97.27  \\ \hline
Combined Dataset & 95.20 & 95.90  & 95.50  & 98.40 \\ \hline
\end{tabular} 
\end{table*}

\begin{figure*}[!t]
     \centering
      \begin{subfigure}[t]{0.49\textwidth}
     \centering
     \includegraphics[width=7.5cm]{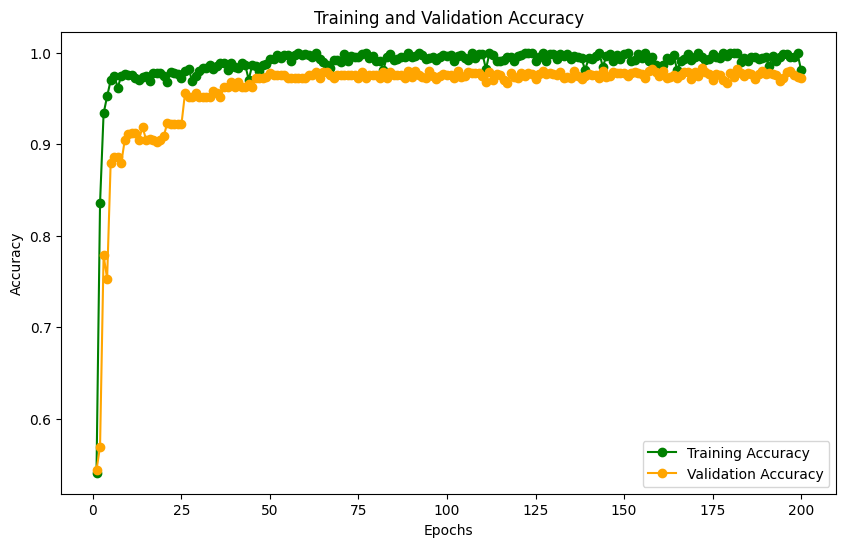}
     \caption{Training and Validation Accuracy curve (binary-class)}
    \label{fig:Training_Accuracy_Dataset-2}
     \end{subfigure}
    \hfill
    \begin{subfigure}[t]{0.49\textwidth}
         \centering
         \includegraphics[width=9cm]{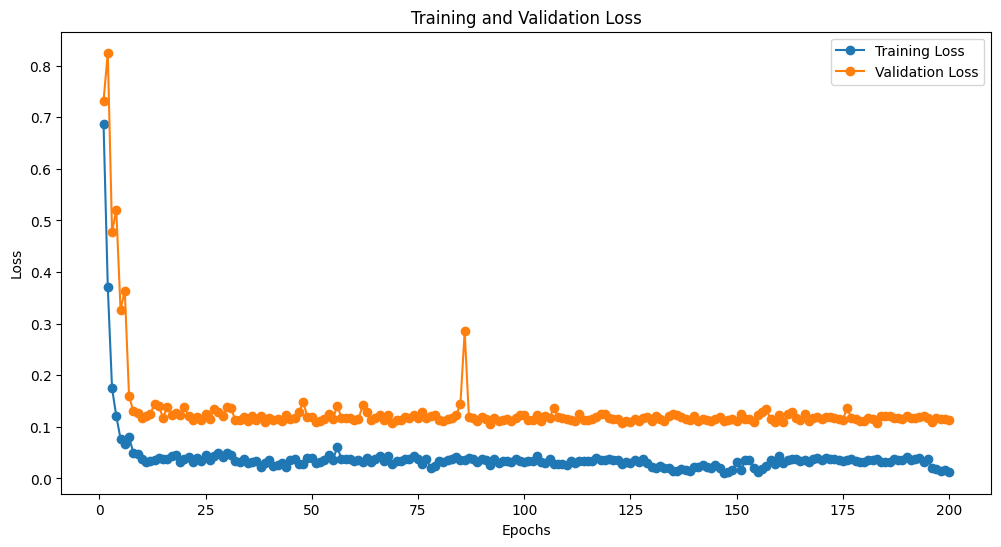}
         \caption{Training and Validation Loss curve (binary-class)}
         \label{fig:Training_loss_Dataset-2}
     \end{subfigure}
     \begin{subfigure}[t]{0.49\textwidth}
     \centering
     \includegraphics[width=7.5cm]{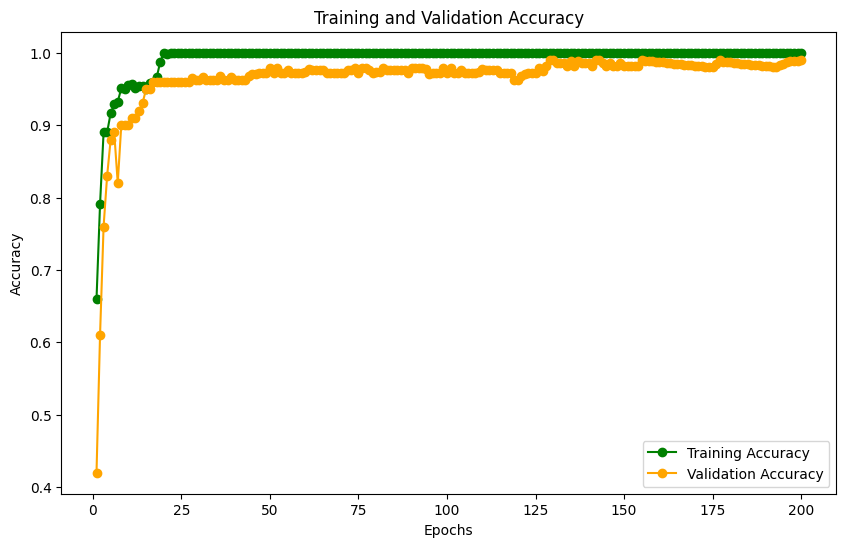}
     \caption{Training and Validation Accuracy curve (multi-class)}
    \label{fig:mendely_multiclasss_accuracy_curve}
     \end{subfigure}
    \hfill
    \begin{subfigure}[t]{0.49\textwidth}
         \centering
         \includegraphics[width=7.5cm]{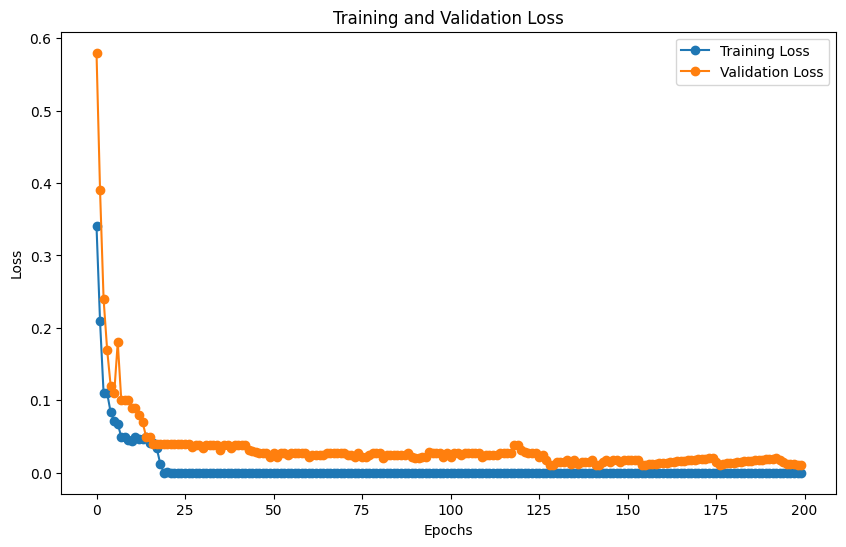}
         \caption{Training and Validation Loss curve (multi-class)}
          \label{fig:mendely_multiclasss_loss_curve}   
     \end{subfigure}
     \caption{(a) Training and Validation Accuracy curve (binary-class), (b) Training and Validation Loss curve (binary-class), (c) Training and Validation Accuracy curve (multi-class), and (d) Training and Validation Loss curve (multi-class) for KXO-Mendeley Multi-class Dataset. }
      
\end{figure*}

%%%%Dataset - mendely_multiclass
Figure \ref{fig:mendely_multiclasss_Confusion_matrix_Dataset}, \ref{fig:mendely_multiclasss_accuracy_curve}, \ref{ig:mendely_multiclasss_loss_curve} demonstrated the confusion matrix, accuracy curve and loss curve of the  KXO-Mendeley Multi-Class dataset for the proposed model. The provided figures illustrate the performance of a classification model trained for a multiclass problem involving three classes: Healthy, Osteopenia, and Osteoporosis. The confusion matrix reveals that the model performs well, with the highest number of correct predictions (29) for the Osteopenia class, followed by 11 for Osteoporosis and 6 for Healthy. Misclassifications are minimal, with only a few samples misclassified between neighboring classes, suggesting good overall accuracy and specificity. The training and validation accuracy plot shows a rapid improvement in accuracy during the initial epochs, stabilizing near 1.0 for both training and validation accuracy, indicating excellent generalization and minimal overfitting. Similarly, the training and validation loss plot demonstrates a sharp reduction in loss during the early epochs, with both metrics converging to low values, further confirming effective model training. Together, these results indicate that the classifier achieves high accuracy, good convergence, and reliable predictions across all three classes.

\begin{figure*}[!t]
     \centering
      \begin{subfigure}[t]{0.49\textwidth}
     \centering
     \includegraphics[width=7.5cm]{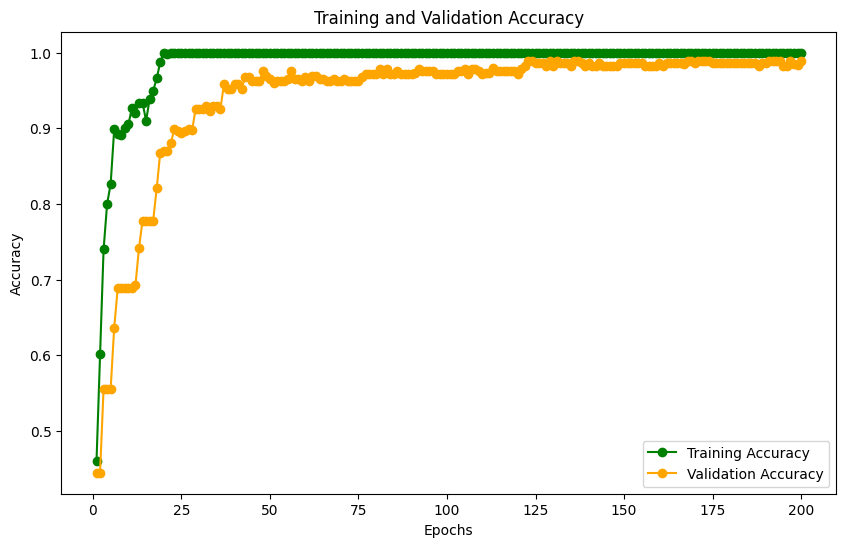}
     \caption{Training and Validation Accuracy curve (binary-class)}
    \label{fig:Training_Accuracy_Dataset-3}
     \end{subfigure}
    \hfill
    \begin{subfigure}[t]{0.49\textwidth}
         \centering
         \includegraphics[width=9cm]{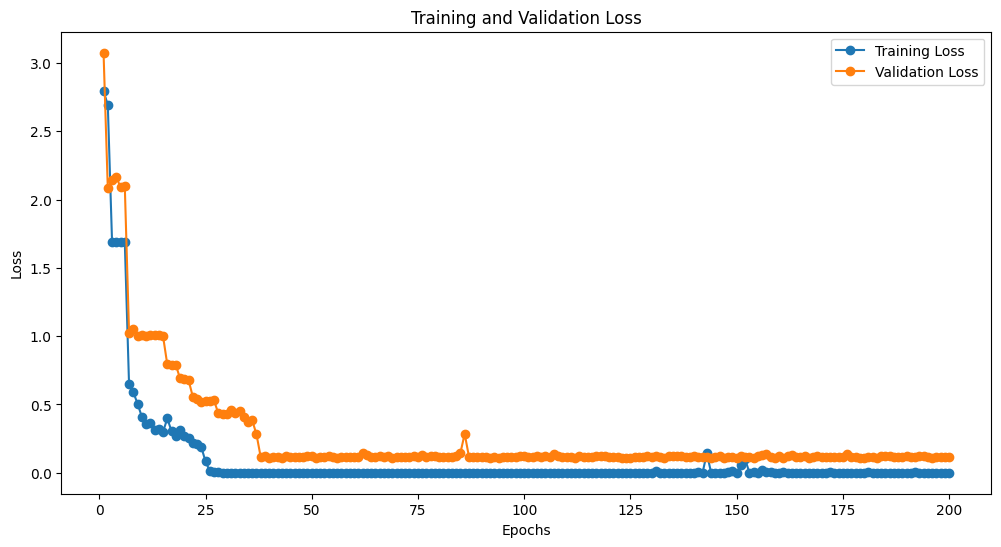}
         \caption{Training and Validation Loss curve (binary-class)}
         \label{fig:Training_loss_Dataset-3}
     \end{subfigure}
     \begin{subfigure}[t]{0.49\textwidth}
     \centering
     \includegraphics[width=7.5cm]{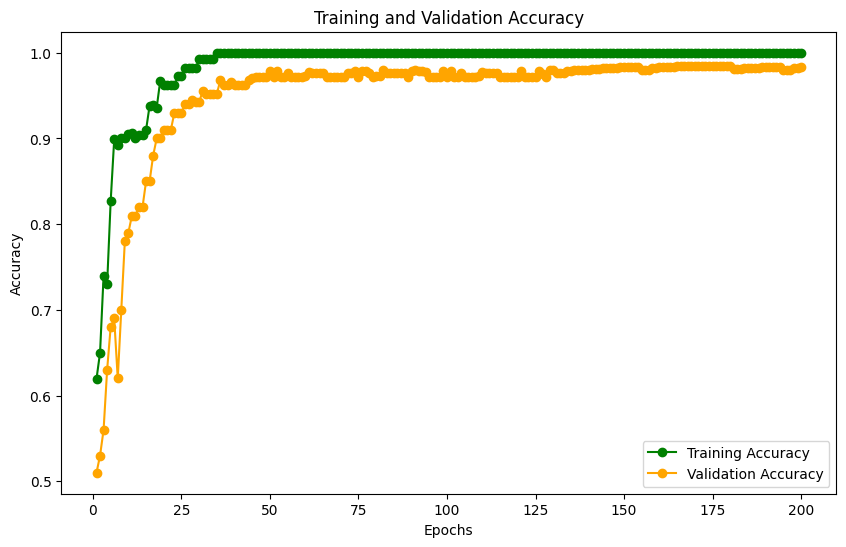}
     \caption{Training and Validation Accuracy curve (multi-class)}
    \label{fig: Kaggle Multiclass_accuracy_curve}
     \end{subfigure}
    \hfill
    \begin{subfigure}[t]{0.49\textwidth}
         \centering
         \includegraphics[width=7.5cm]{Images/Dataset-2_Multiclass_Training_and_validation_loss.png}
         \caption{Training and Validation Loss curve (multi-class)}
         \label{fig: Kaggle Multiclass_loss_curve}
     \end{subfigure}
     \caption{(a) Training and Validation Accuracy curve (binary-class), (b) Training and Validation Loss curve (binary-class), (c) Training and Validation Accuracy curve (multi-class), and (d) Training and Validation Loss curve (multi-class) for OKX Kaggle Multiclass Dataset }
     \label{fig:Kaggle Multiclass_curve}     
\end{figure*}

%%%%Dataset - 3
Besides this, Figure \ref{fig: Kaggle Multiclass_Confusion_matrix_Dataset}, \ref{fig: Kaggle Multiclass_accuracy_curve}, \ref{fig: Kaggle Multiclass_loss_curve} demonstrated the confusion matrix, accuracy curve and loss curve of the OKX Kaggle Multiclass dataset for the proposed model. The confusion matrix indicates that the model effectively classifies 154 samples as Healthy, 72 as Osteopenia, and 155 as Osteoporosis, demonstrating strong precision and recall, particularly for the Healthy and Osteoporosis classes. The accuracy curve shows training accuracy quickly reaching 100\% and validation accuracy stabilizing at around 97\%, suggesting effective learning and strong generalization to unseen data, with minimal overfitting. The loss plot reveals a steady decrease in training loss converging near 0, while validation loss stabilizes just above 0, confirming the model's reliability and efficiency in multiclass classification of knee X-ray images.\\

%%%%%%%Combined Dataset%%%%%%%%%%%%

\begin{figure*}[!t]
     \centering
      \begin{subfigure}[t]{0.49\textwidth}
     \centering
     \includegraphics[width=7.5cm]{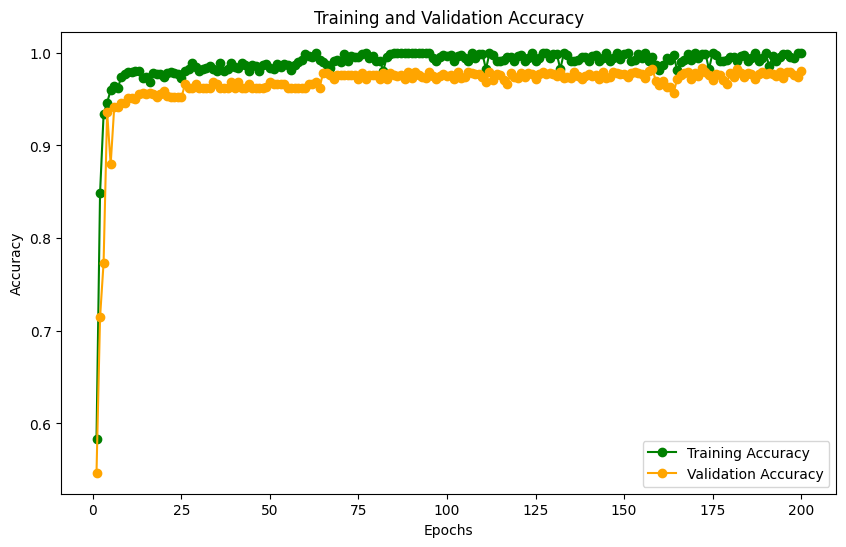}
     \caption{Training and Validation Accuracy curve (binary-class)}
    \label{fig:Combined_Training and Validation Accuracy}
     \end{subfigure}
    \hfill
    \begin{subfigure}[t]{0.49\textwidth}
         \centering
         \includegraphics[width=9cm]{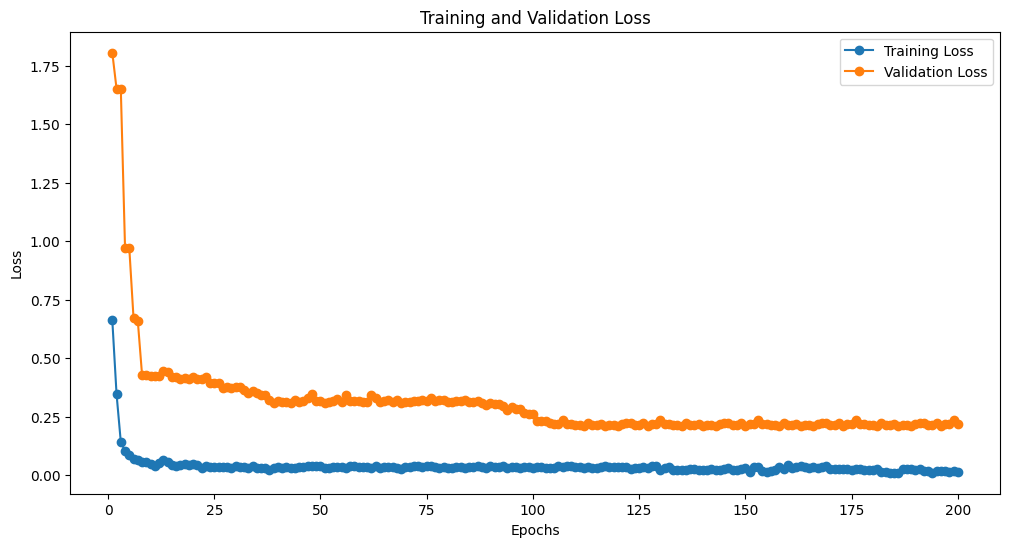}
         \caption{Training and Validation Loss curve (binary-class)}
         \label{fig:Combined_Training and Validation Loss}
     \end{subfigure}
     \begin{subfigure}[t]{0.49\textwidth}
     \centering
     \includegraphics[width=7.5cm]{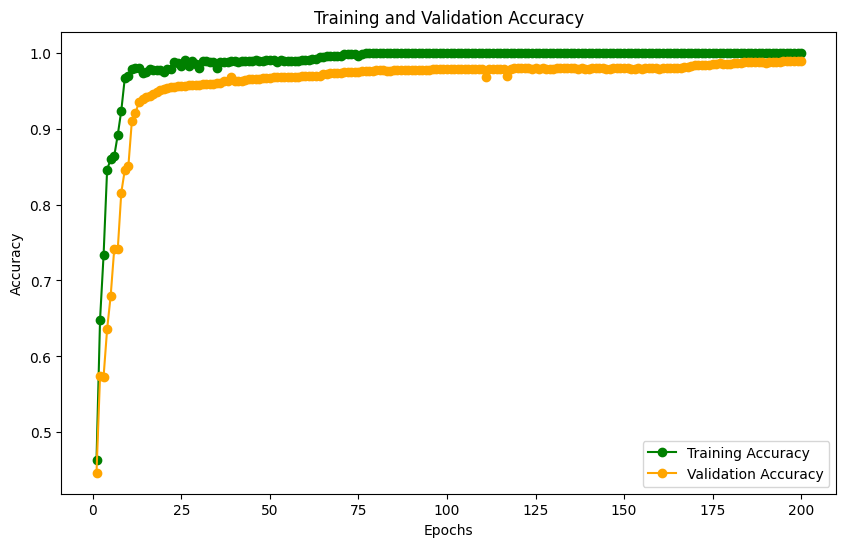}
     \caption{Training and Validation Accuracy curve (multi-class)}
    \label{fig:combined_multiclass_accuracy_curve}
     \end{subfigure}
    \hfill
    \begin{subfigure}[t]{0.49\textwidth}
         \centering
         \includegraphics[width=9cm]{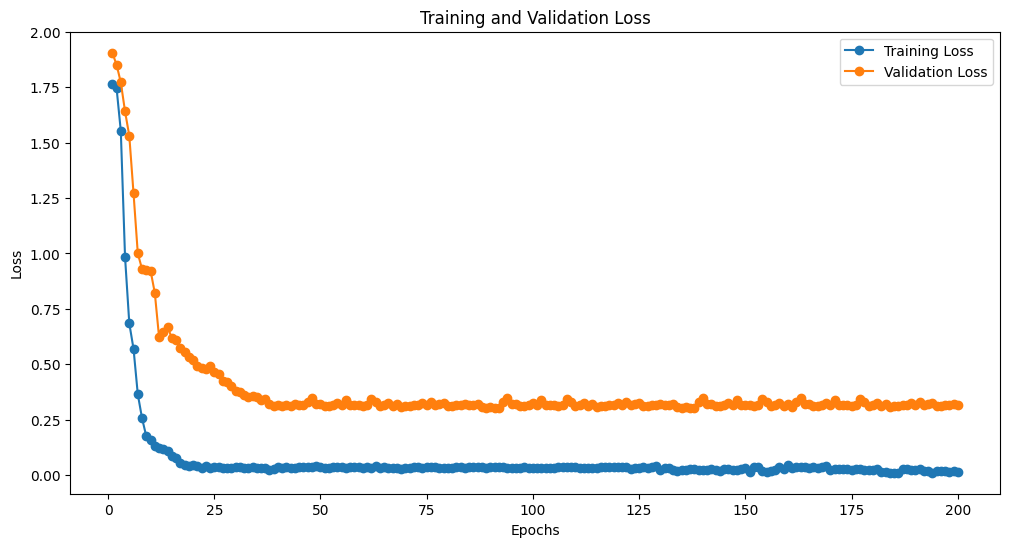}
         \caption{Training and Validation Loss curve (multi-class)}
         \label{fig:combined_multiclass_loss_curve}
     \end{subfigure}
     \caption{ (a) Training and Validation Accuracy curve (binary-class), (b) Training and Validation Loss curve (binary-class), (c) Training and Validation Accuracy curve (multi-class), and (d) Training and Validation Loss curve (multi-class) for Combined Dataset}
     \label{fig:Combined Multiclass_curve}     
\end{figure*}

Figure \ref{fig:combined_multiclass_Confusion_matrix_Dataset}, \ref{fig:combined_multiclass_accuracy_curve}, \ref{fig:combined_multiclass_loss_curve} demonstrated the confusion matrix, accuracy curve and loss curve of the Combined Multi-Class dataset for the proposed model. The confusion matrix in figure \ref{fig:combined_multiclass_Confusion_matrix_Dataset}  highlights the model's strong performance across three classes: Healthy, Osteopenia, and Osteoporosis, with high diagonal dominance indicating excellent classification accuracy. Healthy and Osteoporosis classes show outstanding results with 195 and 203 correct predictions, respectively, and minimal misclassifications. While Osteopenia has slightly higher misclassifications, it still achieves commendable accuracy. Overall, the low off-diagonal values confirm the model’s high sensitivity and specificity, demonstrating reliable performance across all classes. The training and validation accuracy curve \ref{fig:combined_multiclass_accuracy_curve} shows that training accuracy quickly reaches about 100\% by the 50-epoch mark, while validation accuracy stabilizes around 98\%. As training continues, both metrics level off, indicating the model has converged and optimized performance, with minimal gains beyond 50 to 75 epochs. Overall, the results highlight the robustness of the model in classifying knee X-ray images. Figure \ref{fig:combined_multiclass_loss_curve} illustrates the loss trends for training and validation datasets over 200 epochs, highlighting the model's learning efficiency. The training loss decreases steadily to near-zero, demonstrating the model's excellent fit to the training data. The validation loss stabilizes at a low value, indicating strong generalization and minimal overfitting, ensuring reliable performance on unseen data.

\subsection{State of Art Comparison for OKX Kaggle Binary
Dataset}
The state-of-the-art comparison for the OKX Kaggle Binary Dataset, summarized in Table \ref{tab: State of Art Comparison for Kaggle Dataset 1}, highlights the limitations of existing models and the advantages of the proposed approach.

\label{sec: State of Art Comparison}
\begin{table*}[!htp] 
\centering
\caption{State of Art Comparison for OKX Kaggle Binary Dataset} 
\label{tab: State of Art Comparison for Kaggle Dataset 1} 
\begin{tabular}{|l|l|l|l|l|} 
\hline  %\begin{tabular}[c]{@{}l@{}} Dataset \\ Name\end{tabular} 
\textbf{Model Name}  & \textbf{Dataset} & \textbf{Precision(\%)} &  \textbf{Accuracy(\%)} & \begin{tabular}[c]{@{}l@{}} \textbf{Train} \\ \textbf{Test Split}\end{tabular}  \\ \hline  
Ensemble Learning Model(KONet)\cite{ rasool2024konet} & Kaggle Dataset  &  97  & 97& -\\ \hline 
Superfluity DL\cite{naguib2024new} & Kaggle Dataset &  59  & 74.51&   \begin{tabular}[c]{@{}l@{}@{}} Training = 70\%, \\ Validation = 10\%,\\Testing = 20\% \end{tabular}\\ \hline 
GoogLeNet\cite{abubakar2022transfer} & Kaggle Dataset &  59  & 90& \begin{tabular}[c]{@{}l@{}} Training = 80\%, \\Testing = 20\% \end{tabular}\\ \hline 
Proposed Model & Kaggle Dataset &   97.20 &  97.32 &  \begin{tabular}[c]{@{}l@{}l@{}} Training = 60\%, \\Testing = 20\% \\ Validation = 20\% \end{tabular}  \\ \hline
\end{tabular} 
\end{table*}

The KONet ensemble learning model achieved a precision and accuracy of 97\%, showcasing strong performance. However, it lacks detailed metrics such as computational time and loss, making it challenging to evaluate its efficiency fully. The Superfluity DL model, though innovative, underperformed with only 59\% precision and 74.51\% accuracy, indicating significant limitations in handling the dataset effectively. Additionally, the training-validation-testing split (70\%-10\%-20\%) used in this model may not have been optimal for achieving robust performance. Similarly, GoogLeNet, a transfer learning-based model, achieved an accuracy of 90\% but suffered from low precision (59\%), highlighting inefficiency in resource utilization.

In contrast, the proposed model achieves a superior accuracy of 97.32\%, demonstrating its robustness and precision in handling binary classification tasks. Its advanced feature extraction and enhancement using transfer learning and stacked deep learning blocks enable better generalization and higher efficiency. Furthermore, the model addresses drawbacks of existing systems, such as reliance on handcrafted features, inefficiency, and lack of adaptability, making it a more effective solution for osteoporosis diagnosis from X-ray images.

\begin{table*}[!htp] 
\centering
\caption{State of Art Comparison for KXO-Mendely Multiclass
Dataset} 
\label{tab: State of Art Comparison for Mendeley Dataset} 
\begin{tabular}{|l|l|l|l|l|} 
\hline  %\begin{tabular}[c]{@{}l@{}} Dataset \\ Name\end{tabular} 
\textbf{Model Name}  & \textbf{Dataset} & \textbf{Precision} &  \textbf{Accuracy} &  \begin{tabular}[c]{@{}l@{}} \textbf{Train} \\ \textbf{Test Split}\end{tabular}  \\ \hline 
 Superfluity DL\cite{naguib2024new} & Mendeley Dataset  &  73.67  & 83.74 & \begin{tabular}[c]{@{}l@{}@{}} Training = 70\%, \\ Validation = 10\%,\\Testing = 20\% \end{tabular}    \\ \hline
Fuzzy Rank based Ensemble Model \cite{kumar2023fuzzy}& Mendeley Dataset & -   &  93.5 &- \\ \hline
GLCM+FBLS\cite{chen2023glcm}& Mendeley Dataset &  -  &  87.09 &  \begin{tabular}[c]{@{}l@{}} Training = 60\%, \\Testing = 40\% \end{tabular}\\ \hline
Proposed Model & Mendeley Dataset &  97.80  &  98.24 &   \begin{tabular}[c]{@{}l@{}l@{}} Training = 60\%, \\Testing = 20\% \\ Validation = 20\% \end{tabular}\\ \hline
\end{tabular} 
\end{table*}
\begin{table*}[!htp] 
\centering
\caption{State of Art Comparison for OKX Kaggle Multiclass
Dataset} 
\label{tab: State of Art Comparison for Kaggle Dataset 2} 
\begin{tabular}{|l|l|l|l|l|} 
\hline  %\begin{tabular}[c]{@{}l@{}} Dataset \\ Name\end{tabular} 
\textbf{Model Name}  & \textbf{Dataset} & \textbf{Precision} &  \textbf{Accuracy} & \begin{tabular}[c]{@{}l@{}} \textbf{Train} \\ \textbf{Test Split}\end{tabular}  \\ \hline  
 VGG-19\cite{sarhan2024knee} & Kaggle Dataset &  98  & 97.5 &\begin{tabular}[c]{@{}l@{}} Training = 80\%, \\Testing = 20\% \end{tabular} \\ \hline 
Proposed Model & Kaggle Dataset &  93.00  &  97.27&  \begin{tabular}[c]{@{}l@{}l@{}} Training = 60\%, \\Testing = 20\% \\ Validation = 20\% \end{tabular}  \\ \hline
\end{tabular} 
\end{table*}
\subsection{State of Art Comparison for KXO-Mendely Multiclass
Dataset}
\label{sec: State of Art Comparison}
The state-of-the-art comparison for the KXO-Mendeley Multiclass Dataset, summarized in Table \ref{tab: State of Art Comparison for Mendeley Dataset}, evaluates models based on their binary classification performance, as existing studies primarily adopted binary classification for this dataset. Despite the dataset’s multi-class nature, converting it to binary enables direct comparison with existing approaches.
The Superfluity DL model achieved a precision of 73.67\% and an accuracy of 83.74\%, which indicates moderate performance but suggests potential limitations in feature extraction and classification robustness. Its training-validation-testing split of 70\%-10\%-20\% may have contributed to this outcome by limiting the dataset exposure during training.
The Fuzzy Rank-based Ensemble Model demonstrated improved accuracy at 93.5\%, with a reported loss of 0.082. However, the absence of precision or computational time metrics makes it challenging to evaluate its overall efficiency and effectiveness comprehensively.
The GLCM+FBLS model achieved an accuracy of 87.09\% with a loss of 0.082, using a 60\%-40\% train-test split. While promising, its relatively lower accuracy compared to the ensemble model highlights limitations in handling binary tasks derived from multi-class datasets.
The proposed model outperformed all existing systems with a precision of 97.80\% and an accuracy of 98.24\%. Its advanced architecture, which combines transfer learning and stacked feature enhancement blocks, enables superior generalization and efficiency. By addressing the limitations of handcrafted features and inefficient processing in prior approaches, the proposed model demonstrates robust performance, setting a new benchmark for binary classification in the KXO-Mendeley dataset.

\subsection{State of Art Comparison for OKX Kaggle Multiclass
Dataset}
\label{sec: State of Art Comparison}
The state-of-the-art comparison for the OKX Kaggle Multiclass Dataset evaluates the performance of the proposed model against existing approaches, as shown in Table \ref{tab: State of Art Comparison for Kaggle Dataset 2}. This comparison aims to demonstrate the efficiency and robustness of the proposed model.
The VGG-19 model achieved an accuracy of 97.5\% with a precision of 98\% and a reported loss of 0.2054. The training and testing split used for this model was 80\%-20\%, which provided sufficient training data but may have limited its generalization to unseen samples. Despite its strong performance, the higher loss value indicates that VGG-19 may still struggle with fine-grained feature extraction, particularly for complex patterns in multi-class datasets.
The proposed model achieved a precision of 93.00\% and an accuracy of 97.27\%. Its architecture leverages transfer learning combined with stacked feature enhancement blocks, enabling superior feature representation and classification capabilities. This approach minimizes loss and improves generalization, making it more robust for handling complex multi-class tasks. Additionally, the proposed model addresses limitations in computational efficiency and feature extraction, ensuring higher accuracy with lower computational overhead.
This comparison highlights the proposed model’s ability to set a new benchmark for multi-class classification on the OKX Kaggle dataset.
%model name, computational complexity, time, training testing splitting
\subsection{Discussion}
The proposed computer-aided diagnosis (CAD) system for knee osteoporosis detection offers several significant advantages over traditional methods. By leveraging transfer learning and stacked feature enhancement blocks, our approach addresses key limitations in existing systems, particularly in terms of accuracy, efficiency, and generalizability.
One of the primary benefits of the proposed system is its ability to automatically extract hierarchical features from X-ray images. Unlike traditional methods that rely on manual feature extraction, which is both time-consuming and prone to human error, our system uses a pre-trained convolutional neural network (CNN) to extract relevant features, minimizing subjective interpretation by radiologists. This approach not only accelerates the diagnostic process but also ensures consistency in feature extraction, leading to more reliable outcomes. The addition of stacked Conv-RELU-MaxPooling blocks further refines the feature representations, enabling the system to detect subtle variations in bone density and joint structure that are indicative of knee osteoporosis.
Another significant advantage is the use of transfer learning, which allows our model to benefit from large-scale pre-trained models, making it particularly effective when working with smaller, clinical datasets. This technique helps overcome the challenge of insufficient annotated data in medical imaging, a common obstacle in developing robust AI-based diagnostic tools. Furthermore, by employing a multi-block feature enhancement strategy, our model is capable of learning more complex patterns and providing more accurate classifications, even with limited data.

In real-world clinical settings, the proposed system can substantially ease the workload of radiologists and orthopedic specialists. Given the increasing demand for diagnostic accuracy and speed in healthcare, this CAD system offers a practical solution by automating the initial stages of knee osteoporosis detection. Radiologists can rely on the system to rapidly assess X-ray images, with the model’s high accuracy ensuring that cases of osteoporosis are not missed, especially in early stages when preventive intervention is most effective. Moreover, the system’s high performance (with accuracy rates ranging from 97.27\% to 98.24\% across different datasets) demonstrates its potential to provide consistent, reliable results. This level of accuracy is crucial for reducing the risk of misdiagnosis and improving patient outcomes. By accurately distinguishing between healthy and osteoporotic knee conditions, the system can help healthcare providers make more informed decisions regarding preventive measures, treatment plans, and patient management strategies. Additionally, the system’s scalability to larger, combined datasets and its ability to work with diverse data sources suggest that it can be adapted to different clinical environments with varying image quality and patient demographics. This adaptability enhances its real-world applicability, allowing healthcare facilities to integrate the system into existing workflows without requiring extensive retraining or model adjustments.

\section{Conclusion}
\label{sec: Conclusion}
This study introduces an innovative computer-aided diagnosis system that leverages transfer learning and stacked feature enhancement deep learning blocks to address the limitations of existing methods for knee osteoporosis detection. By refining feature extraction and enhancing the hierarchical representation of osteoporotic markers, the proposed system effectively mitigates challenges such as reliance on manual feature extraction and subjective interpretation. The integration of Conv-RELU-MaxPooling blocks facilitates the model's ability to learn complex patterns, enabling accurate differentiation between healthy and osteoporotic knee conditions. Experimental results across multiple datasets validate the model's robustness and superior performance accuracy, highlighting its potential as a reliable tool for clinical application. This approach not only improves diagnostic precision but also underscores the significance of combining advanced deep learning techniques with medical imaging to enhance patient outcomes through early and accurate detection.

Future research could focus on optimizing the system for real-time use, expanding datasets to enhance generalizability, and applying advanced techniques like Explainable AI (XAI) to improve transparency and trust in AI predictions. Integrating multimodal data, such as patient history and other diagnostics, could further enhance accuracy and provide a comprehensive assessment of bone health. This study underscores the transformative role of deep learning in medical imaging, advancing osteoporosis diagnosis and supporting healthcare professionals in improving patient outcomes through AI-driven tools.

\bibliographystyle{IEEEtran}
\bibliography{Reference}
\begin{IEEEbiography}[{\includegraphics[width=1in,height=1.25in,clip,keepaspectratio]{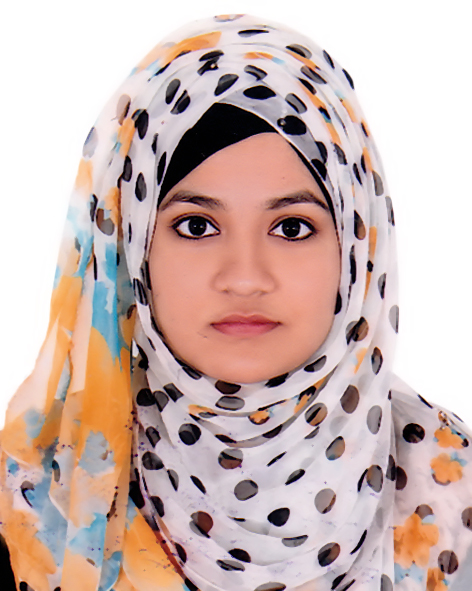}}]{Ayesha Siddiqua} received the B.Sc. degree in computer science and engineering from Bangladesh Army University of Science and Technology (BAUST), Saidpur, Bangladesh in 2021. She is pursuing the M.Sc. degree in computer science and engineering at Bangladesh University of Engineering and Technology, Dhaka, Bangladesh. She is currently serving as a Lecturer at the Department of Computer Science and Engineering, Northern University Bangladesh, Dhaka, Bangladesh. Her research interests include artificial intelligence (AI), machine learning, human-computer interaction (HCI), computer vision, and medical imaging. \end{IEEEbiography}

% \begin{IEEEbiography}[{\includegraphics[width=1in,height=1.25in,clip,keepaspectratio]{Images/Fahmid.jpg}}]{Dr. Fahmid Al Farid} currently working as a post-doctoral scientist at the Faculty of Engineering in Multimedia University. He has completed his Doctor of Philosophy (Ph.D.) by research in information technology from Multimedia University, Cyberjaya, Malaysia. He received a B.S. degree in computer science and engineering from the University of Chittagong, Bangladesh, in 2010 and an M.S. degree from the Faculty of Computer Science and Electrical Engineering, University of Ulsan (UOU), South Korea, in 2015. From 2013 to 2014, he was a Research Assistant in the Embedded System Laboratory at the University of Ulsan, South Korea. In 2015, he was a Research Assistant at the Ubiquitous Computing Technology Research Institute (UTRI), Sungkyunkwan University, South Korea. He received the Korean BK21 PLUS Scholarship, Supported by the Korean Government in MS (2012-2014). He also received an ICT fellowship from Bangladesh Government in 2014. His current research interests include Artificial Intelligence, Algorithm Design, Computer Vision, Human-Computer Interaction, Image, and Video Analysis, Power Generation, and Green Technology.
% \end{IEEEbiography}

\begin{IEEEbiography}[{\includegraphics[width=1in,height=1.25in,clip,keepaspectratio]{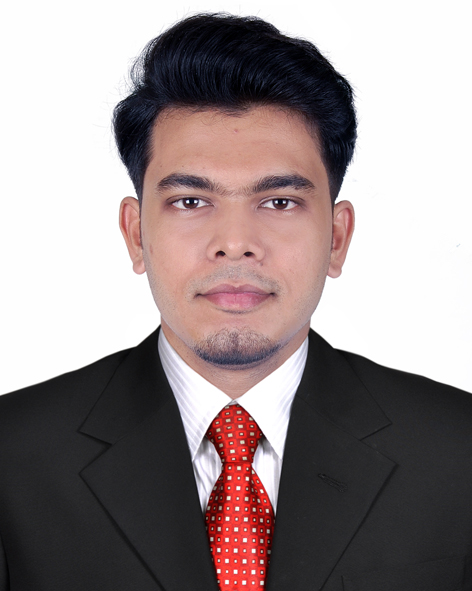}}]{Rakibul Hasan} is an academic professional currently pursuing a Master of Science in Computer Science and Engineering at Brac University. He earned his Bachelor of Science in the same field from Brac University on a full scholarship.
Rakibul has diverse professional experience, including roles as a Lecturer at Northern University Bangladesh and Brac University, and as an SEO Strategist managing content and website optimization. He combines his technical expertise with hands-on experience in software development across various projects.
His research interests include Artificial Intelligence, Machine Learning, Natural Language Processing (NLP), and Data Security. \end{IEEEbiography}

% \begin{IEEEbiography}[{\includegraphics[width=1in,height=1.25in,clip,keepaspectratio]{Images/sarina.jpg}}]{Dr. Sarina Mansor} is a senior lecturer in the Faculty of Engineering at Multimedia University (MMU) Malaysia. She is currently the programme coordinator for B.Eng. (Hons.) Electronics, majoring in Computer. She received her B.Eng. (Hons.) in Electronics and Electrical Engineering from the University of College London in 1998 and her M.EngSc. Degree from Multimedia University in 2002. She completed her DPhil in Engineering Science from the University of Oxford (UK) in 2009. Her research interests are signal and image analysis, medical imaging, computer vision, machine learning \cite{miah2019eeg,kibria2020creation_miah}, and the Internet of Things.
% \end{IEEEbiography}

\begin{IEEEbiography}[{\includegraphics[width=1.10in,height=1.25in,clip,keepaspectratio]{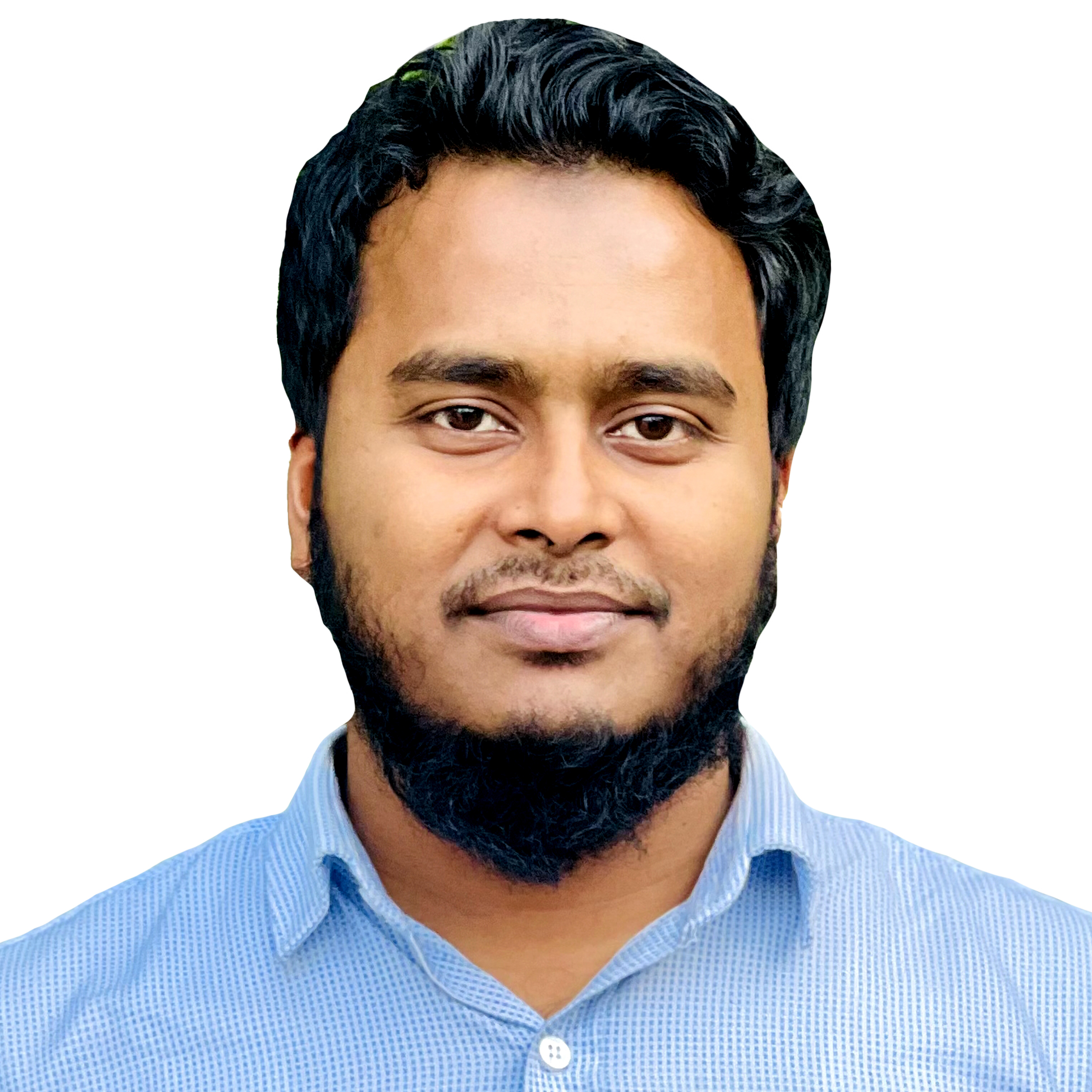}}]{Anichur Rahman} received a B.Sc. and M.Sc degree in Computer Science and Engineering from Mawlana Bhashani Science and Technology University (MBSTU), Tangail, Bangladesh, in 2017 and 2020, respectively. Currently, he is working as an Assistant Professor at Computer Science and Engineering (CSE), National Institute of Textile Engineering and Research (NITER), Constituent Institute of the University of Dhaka, Savar, Dhaka-1350, Bangladesh since January 2020 to present. He has authored more than 40 articles in high-ranking journals and conferences, including FGCS, IEEE IOTJ, IEEE AC, DCAN, JISA, SR (nature), CSBJ, CCJ, etc. Aichur has received the Best Paper Award at the International Conference on Trends in Computational and Cognitive Engineering (TCCE 2020). He has also received the Gold Medal Award from the Bangladesh Textile Mills Association (BTMA) President in Recognition of Excellence in Research. His articles received the best paper award and nominations at different international conferences. He is also a Reviewer of high-quality journals and conferences. His areas of interest include the Internet of Things (IoT), Blockchain, Software Defined Networking (SDN), Network Function Virtualization (NFV), AI, Image Processing, Machine Learning, 5G, Industry 4.0, and Data Science.
\end{IEEEbiography}

\begin{IEEEbiography}[{\includegraphics[width=1in,height=1.25in,clip,keepaspectratio]{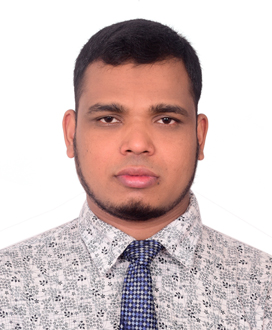}}]{Abu Saleh Musa Miah}  (Member, IEEE) received the B.Sc.Engg. and M.Sc.Engg. degrees in computer science and engineering from the Department of Computer Science and Engineering, University of Rajshahi, Rajshahi-6205, Bangladesh, in 2014 and 2015, respectively, achieving the first merit position. He received his Ph.D. in computer science and engineering from the University of Aizu, Japan, in 2024, under a scholarship from the Japanese government (MEXT). He assumed the positions of Lecturer and Assistant Professor at the Department of Computer Science and Engineering, Bangladesh Army University of Science and Technology (BAUST), Saidpur, Bangladesh, in 2018 and 2021, respectively. Currently, he is working as a visiting researcher (postdoc) at the University of Aizu since April 1, 2024. His research interests include AI, ML, DL, Human Activity Recognition (HCR), Hand Gesture Recognition (HGR), Movement Disorder Detection, Parkinson's Disease (PD), HCI, BCI, and Neurological Disorder Detection. He has authored and co-authored more than 57 publications in widely cited journals and conferences.
\end{IEEEbiography}
\EOD

\end{document}